\documentclass[aps,prx,reprint,superscriptaddress,twocolumn,showkeys,amsmath,amssymb,longbibliography,floatfix]{revtex4-2}
\usepackage[english]{babel}
\usepackage{amsmath,amssymb,bbm,mathrsfs,bm,braket,color,graphicx,comment,amsfonts,dsfont}
\usepackage[colorlinks,linkcolor=blue,citecolor=blue,urlcolor=blue]{hyperref}
\usepackage[mathscr]{euscript}
\usepackage{physics}
\usepackage{xcolor}
\usepackage{ulem}
\usepackage{bm}
\usepackage{orcidlink}
\usepackage{multirow}
\usepackage{lineno}
\usepackage{gensymb}

\def\beq{\begin{equation}}
\def\eeq{\end{equation}}

\begin{document}


\title{
{Beating the aliasing limit with aperiodic monotile arrays}
}

\author{Aurelien Mordret}
\affiliation{Department of Geophysics and Sedimentary Basins, Geological Survey of Denmark and Greenland (GEUS), Øster Voldgade 10, 1350 Copenhagen K, Denmark}
\email{aurmo@geus.dk, corresponding author}
\affiliation{Univ. Grenoble Alpes, Univ. Savoie Mont Blanc, CNRS, IRD, Univ. Gustave Eiffel, ISTerre, 38000 Grenoble, France}

\author{Adolfo G. Grushin}  \affiliation{Univ. Grenoble Alpes, CNRS, Grenoble INP, Institut N\'eel, 38000 Grenoble, France}
\email{adolfo.grushin@neel.cnrs.fr}

\date{\today}

\begin{abstract}
Finding optimal wave sampling methods has far-reaching implications in wave physics, such as seismology, acoustics, and telecommunications. A key challenge is surpassing the Whittaker-Nyquist–Shannon (WNS) aliasing limit, establishing a frequency below which the signal cannot be faithfully reconstructed. However, the WNS limit applies only to periodic sampling, opening the door to bypass aliasing by aperiodic sampling.
In this work, we investigate the efficiency of a recently discovered family of aperiodic monotile tilings, the Hat family, in overcoming the aliasing limit when spatially sampling a wavefield. By analyzing their spectral properties, we show that monotile aperiodic seismic (MAS) arrays, based on a subset of the Hat tiling family, efficiently surpass the WNS sampling limit. Our investigation leads us to propose MAS arrays as a novel design principle for seismic arrays. We show that MAS arrays can outperform regular and other aperiodic arrays in realistic beamforming scenarios using single and distributed sources, including station-position noise and station failures. 
While current seismic arrays optimize beamforming or imaging applications using spiral or regular arrays, MAS arrays can accommodate both, as they share properties with both periodic and aperiodic arrays. More generally, our work suggests that aperiodic monotiles can be an efficient design principle in various fields requiring wave sampling. 

\end{abstract}

\maketitle




\section{Introduction}

Sampling -- the operation of measuring a continuous analog signal at discrete intervals in time, space, or both -- is ubiquitously used in daily life. We find examples in digital audio and video recording and transmission, the Global Positioning System, and health monitoring devices, among many others. It is also an essential component of scientific research, from mathematics and physics to biology, medicine, and Earth science \cite[e.g.,][]{costa2005multiscale, sarvazyan1998shear, morlet1983sampling}.
Sampling relies on
a similar mathematical description across different fields.
For example, the problem of sampling seismic waves using a seismic array -- compact arrangements of seismometers that sample the wavefield at discrete detector sites~\cite{rost2002array, gal2019beamforming}, mathematically resembles that of electronic waves scattering by crystalline atomic sites~\cite{Ashcroft76}.

The similarities between different fields offer a unified mathematical description, but their differences can be leveraged to improve sampling strategies.
Our work is motivated by one such difference: in seismology, seismometers may or may not be arranged periodically, depending on the desired application, while in solid-state crystals, atoms are arranged periodically.
However, the discovery of aperiodic but long-range ordered atomic arrangement in solids, known as quasicrystals~\citep{Shechtman:1984kf, Janssen2008}
portrayed that wave propagation in aperiodic media can
differ significantly from their periodic counterparts~\cite{Collins2017}.
For example, electronic wave functions scattering from quasicrystalline lattices behave differently from typical metals or insulators, resulting in an unusual electrical-conductance scaling with the system size~\citep{Janssen2008, Stadnik1999}.

Numerous quasicrystals have been either grown or found in nature \citep{onoda1988growing, bindi2009natural},
%
and quasicrystalline metamaterials that scatter other types of waves, such as light or sound, also show peculiar wave propagation.
For example, photonic crystals show a fractal hierarchy of band gaps~\citep{Bandres2016} among other phenomena~\citep{Notomi2004, Levi2011, Kraus:2012iqa,schirmann2023physical}, and acoustic metamaterials can reach isotropic effective speeds of sound and isotropic acoustical activity~\citep{Chen2020}.
Characteristic quasicrystalline phenomena and structures can be seen in numerous other platforms, including ultra-cold atomic lattices~\citep{Sbroscia2020}, decorated metallic surfaces~\citep{Collins2017}, polaritonic systems~\citep{Tanese2014}
microwave networks~\citep{Vignolo2016}, acoustic~\citep{Chen2020} and mechanical~\citep{Wang2020b} metamaterials, botanic~\citep{rivier1986botanical} and even architecture~\citep{lu2007decagonal}.

In exploration geophysics, randomized or aperiodic acquisition parameters (source and receiver positions and source timing) are used to optimize seismic surveys to reduce the amount of data while recovering dense datasets \cite{hennenfent2008simply, herrmann2010randomized}. Aperiodicity is starting to be considered in seismic cloaking applications \citep{kumawat2022numerical}. However, the academic seismology community has yet to embrace quasicrystals as a design principle.

The benefit of embracing aperiodicity in seismology lies in that 
aperiodic sampling patterns can be advantageous in signal processing through signal reconstruction, one of the basis of compressive sensing \citep{donoho2006compressed,  mosher2012non, da2018golden,tsingas2023compressive}. 
The Whittaker-Nyquist–Shannon (WNS) sampling theorem~\citep{Whittaker_1915, Nyquist1928, Shannon1949} restricts the faithful reconstruction of a signal from a finite periodic sampling set.
This theorem establishes a bound for the sampling frequency, which must be twice the signal's maximum frequency to reconstruct it faithfully.
The signal reconstruction suffers from aliasing if the sampling happens at a lower frequency than the WNS limit. 
The WNS sampling theorem applies to periodic sampling, and it was recognized that non-periodic arrays can circumvent its bounds \citep{candes2008introduction}   
%
In particular, quasicrystals provide a stable sampling set~\citep{Matei2010}, defined as one which can reconstruct sparse signals accurately from a set of measurements that are fewer in number than the WNS criterion, even in the presence of perturbative noise. 
Hence, finding optimal quasicrystals for wave-sampling purposes remains an essential challenge with far-reaching implications in wave physics such as seismology, acoustics, or telecommunication.

\begin{figure*}
    \begin{center}
    \includegraphics[width=\textwidth]{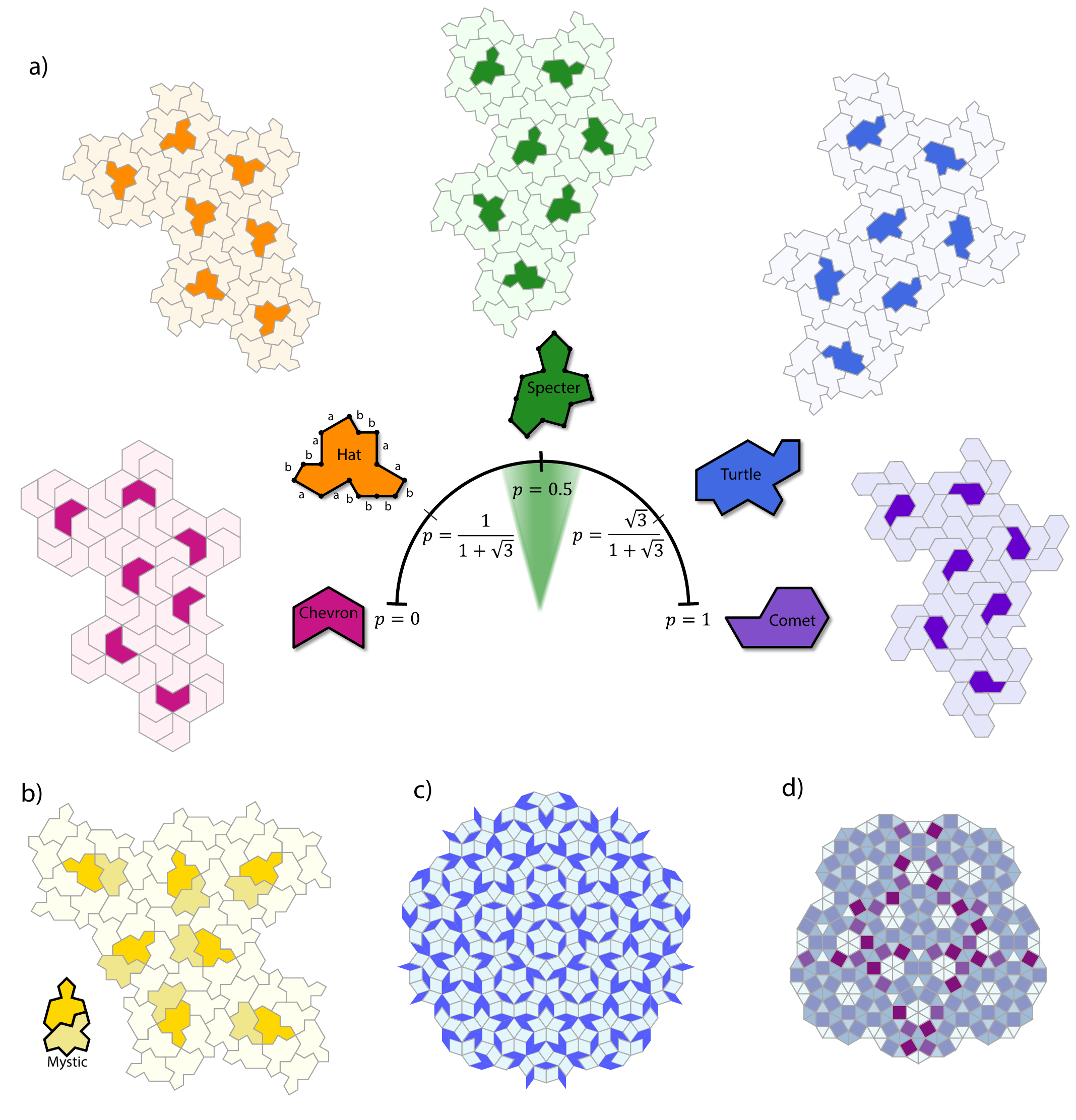} 
    \end{center}
    \caption{%
    \textbf{The Tile$(p)$ family of tilings and other quasicrystals}
    a) We highlight five examples of Tile$(p)$ tilings, from left to right: $p = 0, \frac{1}{1+\sqrt{3}}, 1/2, \frac{\sqrt{3}}{1+\sqrt{3}}, 1$. These particular tilings are formed by tiles whose names are written on the corresponding tiles. Next to each tile, we show a small portion of the tiling where the mirrored tile needed to generate the tiling is outlined in a darker color. The fourteen vertices of each tile are highlighted by black dots for the Hat and the Specter, and the two length scales, $a=p$ and $b=1-p$, are shown around the Hat tile. In the central semicircular diagram, the green shaded area highlights the range of $p$ for which seismic arrays bypass aliasing.  
    b) The Specter tile can also tile the plane aperiodically without using the reflected tile. Instead, it requires the Mystic composite tile, composed of two Specter tiles highlighted with darker colors.
    c) An example of a Penrose tiling using thick and thin rhombi. 
    d) An example of the aperiodic Square-Triangle tiling. 
    }
    \label{fig:HatFamily}
\end{figure*}

Here, we explore a recently discovered family of aperiodic tilings, the Hat family~\citep[Fig.~\ref{fig:HatFamily},][]{Smith2023a, Smith2023b}, to assess its efficiency at beating the WNS sampling limit when sampling spatial seismic wavefields.
We show that, among the whole Hat family of tilings, a parameter region around a member of the family, known as the Specter tiling (see Fig.~\ref{fig:HatFamily}), 
provides ideal properties for unaliased array analysis.
For similar inter-station distance, the Specter outperforms all other members of the Hat tiling, as well as other aperiodic tilings such as the famous Penrose tiling \citep[Fig.~\ref{fig:HatFamily}\textbf{c},][]{penrose1974role, penrose1979pentaplexity} or the Square-Triangle tiling \citep[Fig.~\ref{fig:HatFamily}\textbf{d},][]{oxborrow1993random}.
In contrast, other members of the Hat family whose vertices fall closer to periodic tilings, like the Hat tile, perform worse than those with more irregular vertices coordinates.
Our results suggest that aperiodic monotiles can have significant applications in array seismology specifically and in the rich variety of fields relying on wavefield sampling beyond.

\subsection{Array Seismology}

Historically, array seismology was developed following the United Nations “Conference of Experts to study the methods of detecting violations of a possible agreement on the suspension of nuclear tests” held in 1958 in Geneva \citep{ReportUN}. This event sparked numerous efforts to enhance seismic station capabilities globally. Among these initiatives was the introduction of sensor arrays to improve the signal-to-noise ratio of seismic events, an approach borrowed from radio astronomy, radar, acoustics, and sonar. By the 1960s, it had been proven that seismic arrays outperformed traditional single three-component stations in discriminating seismic signals from both earthquakes and explosions \citep[e.g.,][]{carpenter1965historical, carpenter1965explosion, green1965principles, capon1968short, capon1969long}. 

The deployment of seismic arrays enables the precise detection and analysis of seismic waveforms emanating from various sources, such as tectonic events, volcanic activity, or anthropogenic signals. These arrays, composed of seismometers spatially distributed over a region, allow us to sample the seismic wavefield with a high degree of spatial resolution. The cornerstone of array seismology lies in its ability to employ data processing techniques, notably \textit{beamforming}, to extract and enhance seismic signals of interest from the background noise. Beamforming methods, in general, rely on a delay-and-sum approach. A signal recorded by the elements of an array will stack coherently when the delays due to the array geometry are adequately taken into account, while the incoherent noise will stack destructively, improving the signal-to-noise ratio (SNR) of the signal of interest \citep[see Methods section, Fig.~\ref{fig:BeamformingSUPPL}, ][]{rost2002array, gal2019beamforming}.
This technique is instrumental in understanding the characteristics of the seismic wavefield, facilitating the identification of seismic events, and investigating subsurface structures. 

The recent development of inexpensive autonomous and wireless seismic sensors, called nodes, has boosted the deployment of numerous and relatively dense seismic arrays made of hundreds or thousands of instruments \citep[e.g.,][]{lin2013high, mordret2019shallow}. These so-called large-N arrays, inherited from industrial hydrocarbon exploration, were initially deployed for seismic \textit{imaging} studies, analyzing seismic waves traveling in the subsurface between sources and receivers to generate numerical images of the structures inside the Earth. Later, they were repurposed as detectors of seismic events, using various beamforming or matched-field processing approaches \citep[e.g.,][]{gradon2019analysis,castellanos2020using}. Often, these arrays are laid out on a regular grid, not optimized for array analysis, and suffer from aliasing problems \citep{gradon2019analysis}. On the other hand, arrays designed explicitly for beamforming measurements, such as spiral arrays \citep{kennett2015spiral}, are not ideal for imaging applications as they produce uneven seismic ray (paths joining sources and receivers) densities beneath the array. Therefore, finding an optimal array geometry that could accommodate beamforming and imaging studies is of great interest to the seismological community.

\subsection{Aperiodic Monotiles}

The field of aperiodic tilings, which describes quasiperiodic arrays of vertices of tiles, was recently shaken by the discovery of a new family of tilings: the Hat family~\citep{Smith2023a, Smith2023b}. 
A tiling is a set of tiles that can tesselate the Euclidean plane without gaps or overlaps.
A set of tiles may tile the plane periodically, aperiodically, or both.
Strong aperiodicity requires that a set of tiles can only tile the plane aperiodically.
The problem of finding an aperiodic monotile, i.e., a unique shape that can tile the plane only aperiodically, concluded in March 2023, after more than half a century of search, when \cite{smith2023aperiodic, smith2023chiral} proposed the Hat family of tilings (Fig.~\ref{fig:HatFamily}).
Unlike other aperiodic tilings, the Hat family of tiles are examples of monotiles that can tile the plane only aperiodically.
They improve on the previous record holder, the Penrose tilings \citep[Fig.~\ref{fig:HatFamily}\textbf{c};][]{penrose1974role, penrose1979pentaplexity}, that required a minimum of two tiles to tile the plane aperiodically.

The aperiodic tiling family can be constructed from a base, 14-sided polygon whose parallel sides have one of two lengths, $a$ or $b$, (see Fig.~\ref{fig:HatFamily}\textbf{a}) 
The sides can be parameterized by a single parameter $p \in [0,1]$ with $a=p$ and $b=1-p$. The Hat tile amounts to choosing $p=\frac{1}{1+\sqrt{3}}$, while the Specter is the more symmetric choice $p=\frac{1}{2}$ giving an equilateral tile. The Turtle tile is found for $p=\frac{\sqrt{3}}{1+\sqrt{3}}$. 
The parameter $p$ interpolates continuously between the Chevron ($p=0$) and the Comet ($p=1$). These are the two limiting cases that, together with the Specter, can also tile the plane periodically [Fig.~\ref{fig:HatFamily}(a)]. 
%
Hereafter, we will refer to the Hat family of tiles as Tile($p$), referring to each tile by its $p$ value.
We will use the vertices of single tiles ($14$ stations) as the basis of single-Tile($p$) MAS arrays.
We will refer to the possible tilings by multiple tiles in the family as Tile($p$) tilings, specified by the $p$ value.
We will use the vertices of such extended tilings (several hundreds of stations) to define large-N arrays.

Within the Hat family, the Specter tile, or Tile$\left(\frac{1}{2}\right)$, is special\footnote{Since we are not interested in strong aperiodicity in this work we use the names Tile$\left(\frac{1}{2}\right)$ and Specter interchangeably. In \cite{Smith2023b}, the Specter has wiggled edges deformed from Tile$\left(\frac{1}{2}\right)$ to enforce strong aperiodicity. For beamforming purposes, only the inter-site distances and relative angles matter. Hence, the Tile$\left(\frac{1}{2}\right)$ and Specter have the same beamforming properties.}.
While Tile$\left(p\neq0,\frac{1}{2},1\right)$ require both the tile and its mirrored image [Fig.~\ref{fig:HatFamily}\textbf{a}] to tile the plane aperiodically, Tile$\left(\frac{1}{2}\right)$ can be made to require only the Specter~\citep{Smith2023b} and not its mirrored image.
To achieve this, one forces the use of the Mystic tile (Fig.~\ref{fig:HatFamily}\textbf{b}), which combines two Specters.
This means that the tiling with Tile$\left(\frac{1}{2}\right)$ has two versions, one without and one with its mirror tile; we call these chiral and achiral Tile$\left(\frac{1}{2}\right)$, respectively.
Both can tile the plane aperiodically and provide equally good performances at beating aliasing. 
For completeness, we note that Tile$\left(\frac{1}{2}\right)$ can also tile the plane periodically, and therefore it is said to be weakly aperiodic~\citep{Smith2023a}, unlike other members of the Hat family.
This latter property does not affect any of our results.


Since its discovery, the novel mathematical properties of the Hat tiling ($p=1/(1+\sqrt{3})$) have motivated a few novel physical results. When considered as a hypothetical two-dimensional material, the Hat tiling's physical properties lie between those of Graphene and those of aperiodic quasicrystals~\citep{schirmann2023physical}.
The particular connectivity of the Specter tiling allowed the authors of Ref.~\cite{Singh24} to find exact solutions for dimer models on this tiling. 
Other authors have considered the Ising model~\cite{Okabe_2024} or the macroscopic elastic behavior of the Hat tiling~\cite{Rieger2023}.
However, previous physically motivated research on monotile tilings focused on either the Hat or the Specter tiles. 
How physical properties change with $p$ remains unexplored.

Each tiling, for any $p$, can be grown from a single tile by following substitution rules~\cite{smith2023aperiodic,smith2023chiral}.
These rules gather a collection of tiles into metatiles, which are then assembled through matching rules.
The procedure can be iterated to create different generations of tilings.
The web-app put forward by the discoverers of the Hat family,
can be used to visualize this procedure \citep{weblink}.
We append with this work the Python codes that reproduce these tilings, created by adapting the JavaScript code of the cited website into Jupyter Notebooks. See the Data and Code availability section for details on accessing these scripts. 

\section{Beamforming properties of single Tile($p$) arrays
\label{sec:singletile}
}

We start by describing the properties of seismic arrays made of seismic stations placed at the vertices of a single Tile($p$), leaving the discussion of seismic arrays based on using multiple tiles to Section~\ref{sec:multipletile}.

We will first analyze the beamforming performances of such arrays by computing their corresponding array response function (ARF). 
The ARF, also known as the structure factor of the lattice, is defined as 
\begin{equation}
    ARF(\mathbf{k}) = \frac{1}{N^2}\left| \sum_{i=1}^N e^{j \mathbf{x}_i \cdot \mathbf{k}}  \right|^2,
    \label{eq:ARFmain}
\end{equation}
where $N$ is the number of stations in the array, $\mathbf{x}_i$ their positions, $\mathbf{k}$ is the wavenumber vector and $j^2=-1$ (see Methods Section for a detailed definition).
In crystallography, the ARF of a solid determines its diffraction pattern~\cite{Ashcroft76,Janssen2008}.
Then, we perform synthetic beamforming for different scenarios corresponding to the illumination of the arrays by a single plane wave coming from a single azimuth with a constant velocity or by multiple plane waves covering all azimuths to mimic a homogeneous distribution of seismic noise sources \citep[e.g., ][]{harmon2010distribution, mordret2013near}. We will compare the robustness against aliasing displayed by Tile($p$) arrays to those of other aperiodic and periodic arrays (see Methods Section and Fig.~\ref{fig:ARFSupplementary} for a self-contained description of aliasing effects). 

\subsection{Array Response Functions of single Tile($p$) arrays}

\begin{figure*}
    \begin{center}
    \includegraphics[width=0.99\textwidth]{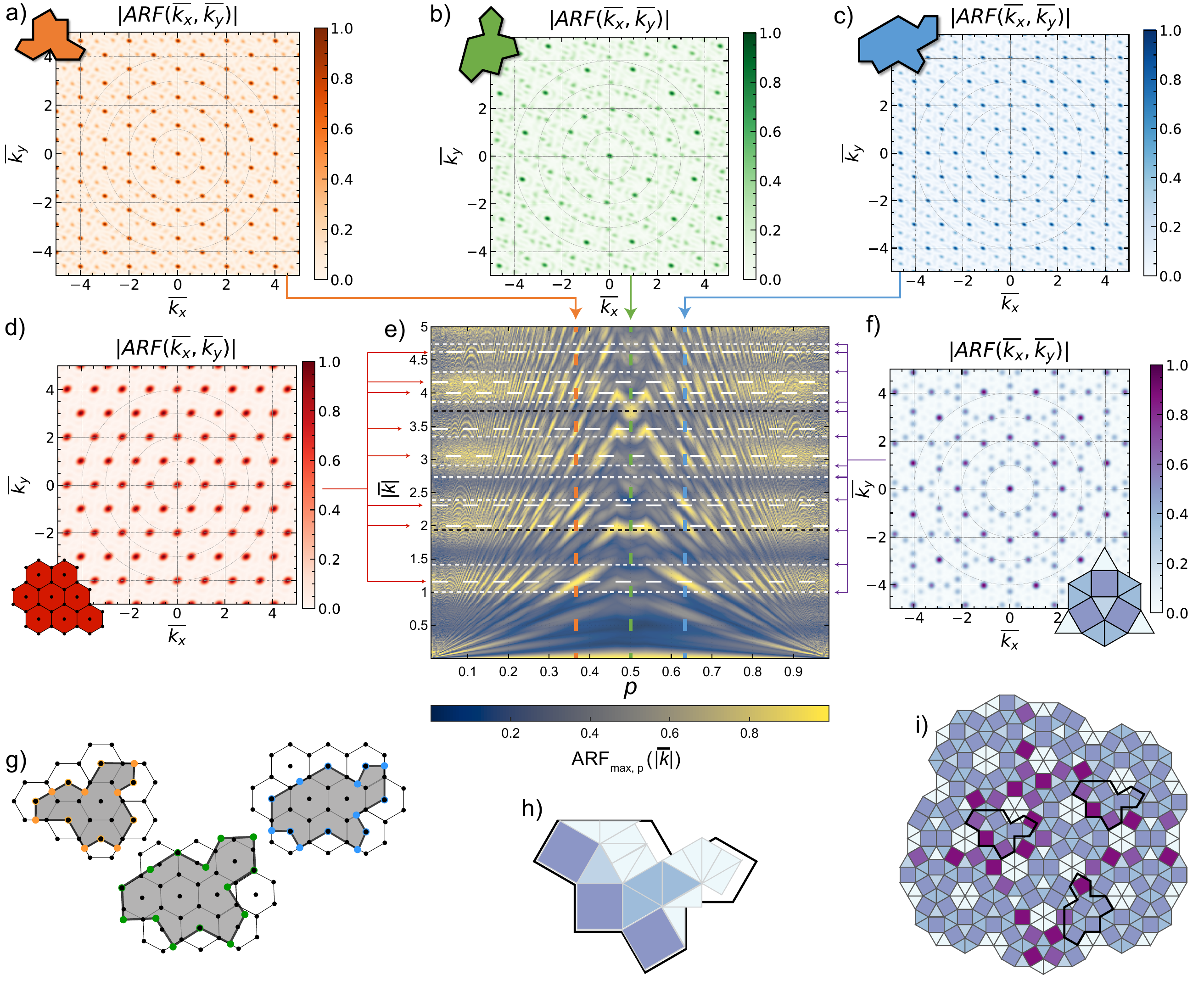} 
    \end{center}
    \caption{\textbf{Single Tile(p) ARF analysis}. a) ARF of the Hat, b) the Specter, and c) the Turtle as a function of the normalized wavenumber $|\mathbf{\overline{k}}|$. d) and f) show the ARF of a Triangular lattice (hexagonal lattice plus its centers) array and the Square-Triangle metatile, respectively. For each $p$, e) shows the peak sidelobe level as a colormap, calculated using Eq.~\eqref{eq:ARFmax} for each $|\mathbf{\overline{k}}|$. The white horizontal dashed lines, indicated by red arrows show the positions of the maxima of the Triangular array ARF (d): $ \left[\frac{2n}{\sqrt{3}}, 2n \right], n \in \mathbb{N}$, and their combinations. The white horizontal dotted lines show the maxima of the Square-Triangle metatile ARF (f), indicated by the purple arrows. The black dotted horizontal lines show the positions of the two main sidelobes in common between the Specter and the Square-Triangle ARFs. g) Illustration of the shared vertices between a Triangular lattice and some Tile($p$) vertices. The open-colored circles show the shared vertices. The solid circles show the Tile($p$) vertices falling outside of the Triangular lattice. h) Re-arrangement of the tiles of the Square-Triangle metatile (g) to form a pseudo-Specter. i) Correspondence between a Square-Triangle aperiodic tiling and the Specter (delineated in black).
    }
    \label{fig:ARFSingleTile}
\end{figure*}

We begin our investigation by analyzing arrays made of a single Tile($p$), i.e., 14 stations (Fig.~\ref{fig:HatFamily}). To compare all arrays on a common scale, we normalize their ARF by the minimum interstation distance $r_{min}$ so that the wavenumber is expressed in terms of multiples of the minimum spatial frequency unit in the array. Effectively, we represent Eq.~\eqref{eq:ARFmain} as a function of a normalized wave-vector $\mathbf{\overline{k}} = \dfrac{r_{min} \cdot \mathbf{k}}{2\pi}$. Figs.~\ref{fig:ARFSingleTile}\textbf{a}, \textbf{b}, and \textbf{c} show the normalized ARFs for the Hat, Specter, and Turtle tiles, respectively. The Hat ARF can be directly compared with the diffraction patterns obtained by \cite{kaplan2024periodic}, who used a much larger set of tiles. 

By construction, the vertices of the Hat monotile are a subset of vertices and center points of a hexagonal or equilateral triangular lattice \citep{smith2023aperiodic}, as exemplified by Fig.~\ref{fig:ARFSingleTile}\textbf{g}. Half of the vertices of the Hat share vertices with the triangular lattice (open colored circles in Fig.~\ref{fig:ARFSingleTile}\textbf{g}). Therefore, it is not surprising that the Hat ARF (Fig.~\ref{fig:ARFSingleTile}) bears a strong resemblance with the ARF of a triangular lattice and its sixfold rotational symmetry~\citep{Socolar2023}, with the brightest spots of the Hat ARF being at the same positions as the ARF of the triangular tiling. The same observation can be made for the Turtle, albeit with a 30$\degree$~rotation of the ARF pattern (Fig.~\ref{fig:ARFSingleTile}\textbf{c}). 

In contrast, the ARF of the Specter tile (Tile$(\frac{1}{2})$, Fig.~\ref{fig:ARFSingleTile}\textbf{b}) exhibits less regularity at small normalized wavenumbers, with no apparent six-fold rotational symmetry and lower amplitude peaks. 
%
Despite the angles between consecutive vertices being the same for all Tile($p$), they can produce very different ARFs.
The Specter's ARF is the only one to exhibit an approximate twelve-fold rotational symmetry.

Recalling the mathematical similarity between seismic array placements and atomic positions in solids, the Specter's ARF resembles structure factors of aperiodic Square-Triangle tillings \citep{oxborrow1993random}, which also exhibit a twelve-fold symmetry (see Fig.~\ref{fig:ARFSingleTile}\textbf{g}). These Square-Triangle tillings can be related to several kinds of quasicrystals realized by twisting graphene atomic monolayers \citep{ahn2018dirac, yao2018quasicrystalline}, or macroscopic spheres arrangements \citep{fayen2024quasicrystal}. The similar symmetry between the Specter's ARF and the Square-Triangle tilling's structure factor suggests that the latter lattice can closely approximate the Specter tile. Similar to the Hat and other Tile($p$) sharing a large number
of its vertices with a triangular lattice, the Specter shares approximately half of its vertices with the Square-Triangle lattice (Fig.~\ref{fig:ARFSingleTile}\textbf{i}). Another way to illustrate this similarity is by approximating the Specter tile (Fig.~\ref{fig:ARFSingleTile}\textbf{h}) with a re-arrangement of the square and triangle metatile shown in (Fig.~\ref{fig:ARFSingleTile}\textbf{i}). The fact that the Specter does not align perfectly with this set of tiles explains the fading of some of the peaks in the ARF compared to the ARF of the Square-Triangle array.

We now wish to assess the performance of Tile($p$)-based arrays regarding aliasing for all $p$ values.
Aliasing is mainly controlled by the position of the most prominent amplitude peaks in the ARF, and thus, we need to locate them in the $[\overline{k_x},\overline{k_y}]$-space. 
To do so, we compute the peak sidelobe level $ARF_{max,p}(|\mathbf{\overline{k}}|)$ \citep{hopperstad98_norsig} as the maximum value of each $ARF_p$ along circles of constant $|\mathbf{\overline{k}}|$ values.
Mathematically, this operation can be expressed as 
\begin{equation}
\label{eq:ARFmax}
        ARF_{max,p}(|\mathbf{\overline{k}}|) = \max_{\theta \in [0, 2\pi]}{ARF_p(|\mathbf{\overline{k}}|\cos{\theta},|\mathbf{\overline{k}}|\sin{\theta})},
\end{equation}
where $|\mathbf{\overline{k}}| = \sqrt{{\overline{k}_x}^2 + {\overline{k}_y}^2}$.
Fig.~\ref{fig:ARFSingleTile}\textbf{e} shows $ARF_{max,p}(|\mathbf{\overline{k}}|)$ for $p \in [0, 1]$ and $|\mathbf{\overline{k}}| \leq 5$. We notice several interesting features. First, higher-intensity background horizontal bands can be seen across almost all $p$. They match the positions of the peaks expected for the triangular lattice, indicated by horizontal dashed lines. Second, this pattern interferes with an M-shape pattern centered at $p=0.5$.
For $p=0.5$, the maxima as a function of $|\mathbf{\overline{k}}|$ fall on some of the Square-Triangle metatile peaks (black horizontal dotted lines in Fig.~\ref{fig:ARFSingleTile}\textbf{e}).
These maxima seem unaffected by the triangular lattice background horizontal band modulations.

Moreover, Tile($p \in [0.45, 0.55]$) does not have any prominent peaks for $|\mathbf{\overline{k}}| \leq 2$, which suggest an enhanced beamforming performance with respect to aliasing compared to regular arrays.
This can be confirmed by noting that the first aliasing peaks of the triangular and square lattice fall at $|\mathbf{\overline{k}}| = \frac{2}{\sqrt{3}} \approx 1.155$ and $|\mathbf{\overline{k}}| = 1$, respectively.
Hence, a monotile aperiodic seismic (MAS) array with station positions at the vertices of a single Tile($p \in [0.45, 0.55]$) should outperform regular arrays for beamforming analysis with respect to aliasing.

\subsection{Beamforming performances of single Tile($p$) arrays}

We now describe synthetic beamforming tests to illustrate the performances of Tile($p$)-based arrays. We use two types of plane-wave source distributions: 
1) distributed sources, with a homogeneous azimuthal distribution of 400 sources randomly spanning 360$\degree$ around the array. An unaliased beamforming result would show a single high and azimuthally constant intensity ring.
2) a single source coming from -45$\degree$N. An unaliased beamforming result would show a single intensity peak. In both cases, we use plane waves with a wavelength equal to twice the smallest inter-station distance in the array (normalized at 1 km) and a slowness of 0.5 s/km, a typical surface wave velocity at 1 Hz. This configuration of minimum inter-station distance and wavelength is at the aliasing limit for a regular square-lattice array, according to the WNS theorem~\citep{Whittaker_1915, Nyquist1928, Shannon1949}.

\begin{figure*}
    \begin{center}
    \includegraphics[width=\textwidth]{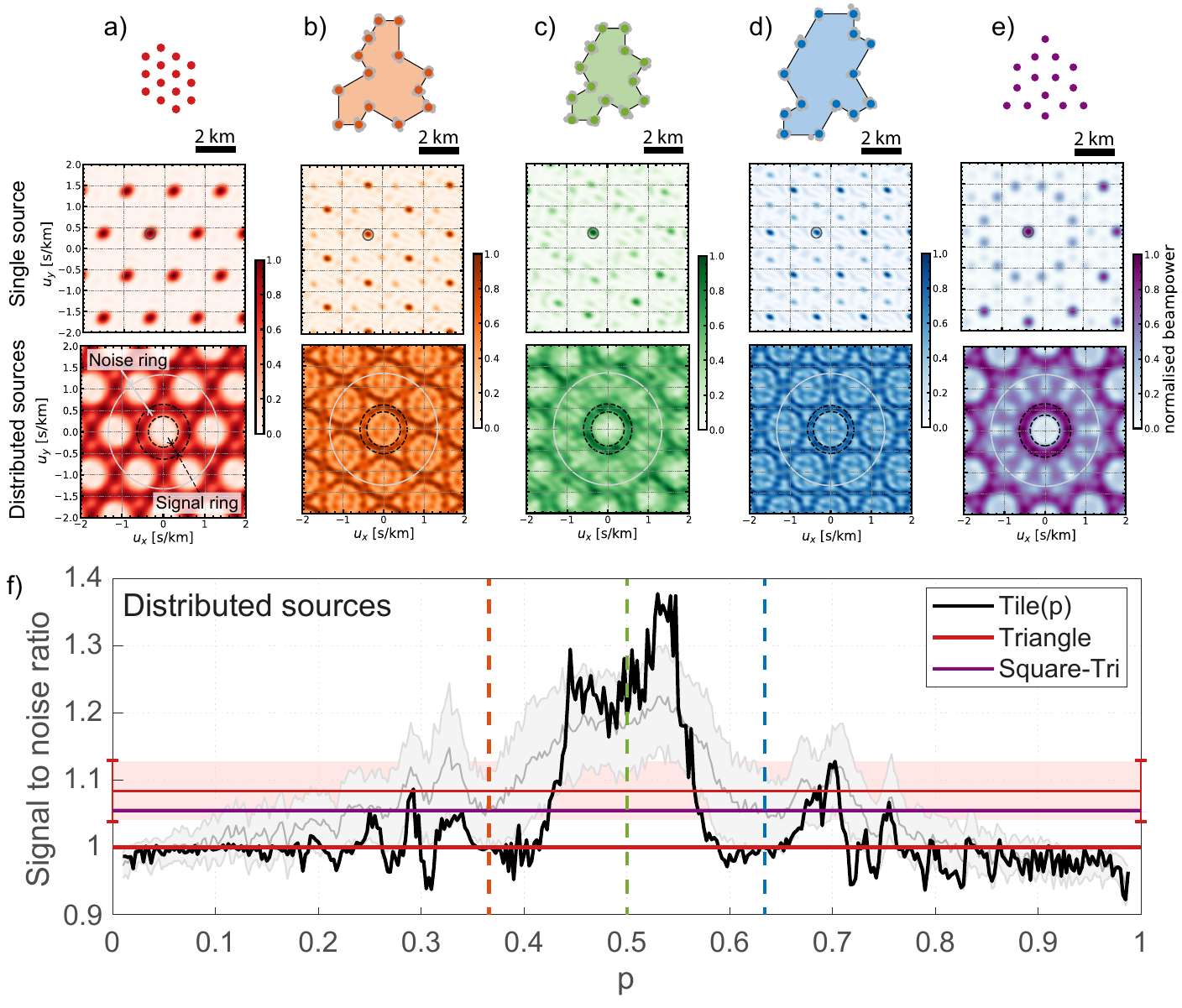} 
    \end{center}
    \caption{
    \textbf{Synthetic beamforming for single Tile($p$) arrays}. The first row shows the maps of the arrays. The grey dots for the Hat, Specter, and Turtle tiles are the randomized positions of the stations. The second row shows the beamforming results for a single source coming from the northwest at 2 km/s, highlighted by the black circle. The third row shows the beamforming results for a homogeneous illumination with sources of 2 km/s plane waves distributed around the arrays. The black dashed circles delineate the ``signal'' ring, and the white circle, the ``noise'' ring used to compute the signal-to-noise ratio (SNR) in f). The different columns show a) a triangular array, b) the Hat, c) the Specter, d) the Turtle, and e) a Square-Triangle metatile. f) SNR of the beamforming results for the distributed sources as a function of the parameter $p$. The gray curve and shaded area show the Tile($p$) SNR's mean and standard deviation, respectively, when perturbing the station positions (gray dots in the first row). Vertical dashed lines highlight the $p$ values of the Hat, Specter, and Turtle arrays. The thick red line shows the SNR for the triangular array. The mean SNR of the triangular array with perturbed positions is shown by the thin red line with the light red shaded area representing the standard deviation. The Square-Triangle array SNR is displayed by the purple horizontal line. The disordered curves are averaged over 100 disorder realizations.
    }
    \label{fig:BeamformingSingleTile}
\end{figure*}

Figure~\ref{fig:BeamformingSingleTile} shows the results of this synthetic beamforming for five arrays: a triangular-lattice array (Fig.~\ref{fig:BeamformingSingleTile}\textbf{a}), the Hat (Fig.~\ref{fig:BeamformingSingleTile}\textbf{b}), the Specter (Fig.~\ref{fig:BeamformingSingleTile}\textbf{c}), the Turtle (Fig.~\ref{fig:BeamformingSingleTile}\textbf{d}), and the Square-Triangle metatile (Fig.~\ref{fig:BeamformingSingleTile}\textbf{e}). For each array, the second row shows the single source beamforming results while the third row shows the distributed sources beamforming results. In agreement with the ARF analysis, it appears that the Hat and the Turtle arrays (Fig.~\ref{fig:BeamformingSingleTile}\textbf{b} and \textbf{d}) behave similarly to the triangular array (Fig.~\ref{fig:BeamformingSingleTile}\textbf{a}) and are strongly aliased. 
In the single source case, Hat and Turtle arrays exhibit multiple high-intensity peaks having the same amplitude as the main peak (highlighted by the black circle in Fig.~\ref{fig:BeamformingSingleTile}). 

In the distributed sources scenario, we also observe a repetition of the main intensity ring (the ``signal'' ring, delimited by the black dashed circles in the third row of Fig.~\ref{fig:BeamformingSingleTile}) across the beamforming diagram. 
These side-rings are tangent to the signal ring. For regular-lattice arrays, the side rings overlap with each other and with the signal ring (Fig.~\ref{fig:BeamformingSingleTile}\textbf{a}), and the maxima of the beamforming are not located at the correct slowness values. 
In a blind beamforming experiment with unknown plane-wave source parameters, extracting meaningful information using the Hat, the Turtle, or the triangular arrays would be challenging because there is no way to identify the main peak (or ring) from its side peaks (or rings) and correctly identify the incoming wave's slowness.  

Notably, the Specter array (Fig.~\ref{fig:BeamformingSingleTile}\textbf{c}) outperforms the other geometries. In the distributed sources configuration, the Specter array, and to a lesser extent, the Square-Triangle array, presents a well-defined signal ring with minimal side-lobes intensity up to $1.33$ s/km. The Specter array shows a homogeneous low-intensity level between 0.5 and 1.33 s/km for the distributed sources, while the Square-Triangle array shows a more pronounced variability. In the single source configuration, the Specter array provides a clear maximum at the expected slowness position of the incoming wave. The side lobes are far from the main peak compared to the Hat, Turtle, and triangular arrays and have smaller amplitudes. The Square-Triangle array has a similar beamforming result, albeit with more numerous and more prominent amplitude side lobes. 

To quantify the performances of the Tile($p$)--based arrays, we compute the signal-to-noise ratio (SNR) of the beamforming results for the distributed sources configuration, using the black dashed rings depicted in the third row of Fig.~\ref{fig:BeamformingSingleTile}. We define the SNR as the ratio of the maximum amplitude of the inner ring over the maximum amplitude of the outer ring. We chose this definition for the SNR because it gives a value of $\approx1$ for the aliased arrays.  
The two rings are defined by constant-slowness radii. The inner ring (the ``signal'' ring) is defined between $u_{in,\pm} = u_0 \pm \frac{1}{2 f_0 r_{max}}$, and the outer ring (the ``noise'' ring) is defined between $u_{in,+}$ and  $u_{out} = 1.33$ s/km, where $f_0 = 1$ Hz, $u_0 = 0.5$ s/km and $r_{max}$ is the maximum inter-station distance or aperture of each array to account for the varying resolution. 

The SNR results in Fig.~\ref{fig:BeamformingSingleTile}\textbf{f}) compares the Tile($p$) in black, the triangular array in red, and the Square-Triangle array in purple. The Tile($p$) family performs at least 20\% better than regular-lattice arrays in the interval $p \in [0.45, 0.55]$. The Square-Triangle metatile performs on average only 5\% better than aliased arrays. While not the absolute best in the range $p \in [0.45, 0.55]$, the Specter tile is a good choice as a compromise between performance and design simplicity for a simple 14-station array, beating the aliasing limit of regular-lattice arrays.

\begin{center}
    \begin{figure}
    \begin{center}
    \includegraphics[width=\linewidth]{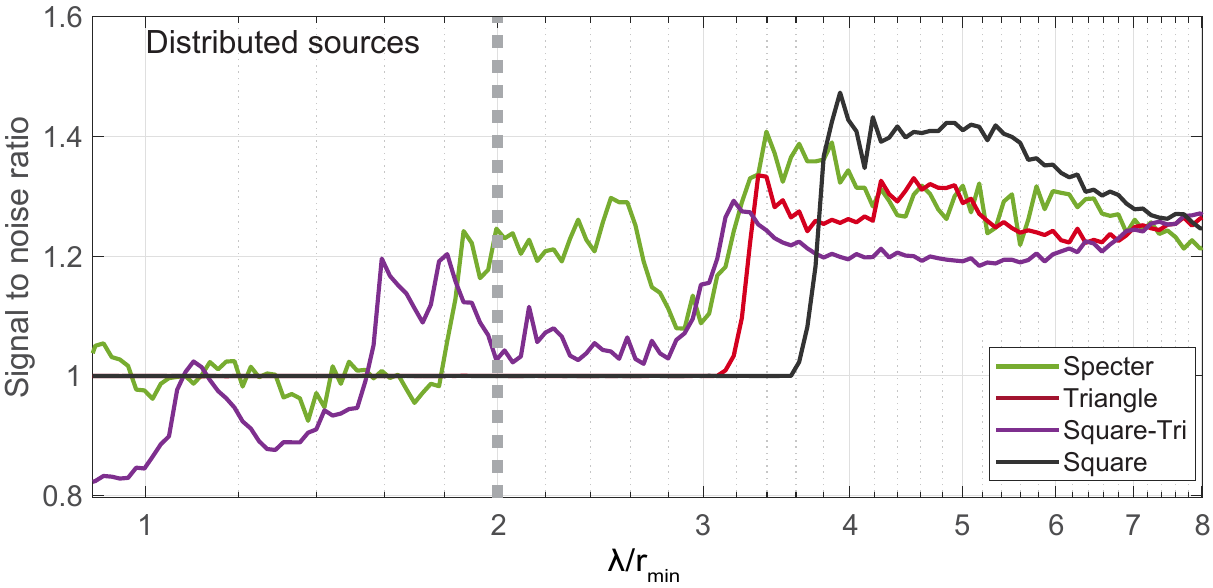} 
    \end{center}
    \caption{
    \textbf{Beamforming SNR as a function of the wavelength}. Beamforming SNR for distributed sources as a function of the normalized wavelength for regular square and triangular arrays and for the Specter and Square-Triangle aperiodic arrays. The vertical dashed line indicates the aliasing limit: when the wavelength $\lambda$ equals twice the minimum interstation distance $r_{\mathrm{min}}$.
    }
    \label{fig:SNRvsLambda}
\end{figure}
\end{center}


We tested the robustness of these results with respect to disordering the position of each station. If a small position error degrades the beamforming performances, then it becomes difficult to envision a field deployment using these geometries. Any obstacle or difficulty in accurately positioning sensors would be detrimental to the quality of the measurements. For each Tile($p$), we varied the position of the 14 sensors randomly drawing from a normal distribution centered on the actual station position and a 100 m variance (10\% of the minimum inter-station distance) for a hundred realizations. These positions are displayed as gray points in the array maps of the first row of Fig.~\ref{fig:BeamformingSingleTile}. The distribution of SNR for each $p$ is shown as the dark gray curve with the gray contour for its mean and standard deviation, respectively. We see that, for $p \notin [0.45, 0.55]$, perturbing the station positions improves the SNR, whereas, for $p \in [0.45, 0.55]$, the SNR is slightly reduced but still exhibits a local maximum. In comparison, perturbing the positions of the triangular array also improves the SNR by about 10\%. Still, it remains below the level of Tile($p$), and on par with the Square-Triangle arrays SNR level.

The above analysis was calculated at the aliasing limit when the wavelength equals twice the minimum interstation distance in the array.
We now explore whether the Specter array performs better than the other arrays as we vary the wavelength. Figure~\ref{fig:SNRvsLambda} shows how the beamforming SNR varies when varying the normalized wavelength, $\bar{\lambda}=\lambda/r_{min}$. The different curves represent the Specter (green), the triangular (red), and the Square-Triangle (purple) arrays, used in Figure~\ref{fig:BeamformingSingleTile}, and a square-lattice regular array (black). The four arrays exhibit a sharp increase of SNR for $\bar{\lambda}\approx 3.5$, a value for which the aliased ring patterns are pushed beyond the slowness limits of the 'noise ring'. Whether this value is general for most seismic arrays needs further analysis beyond the scope of this paper, but if this is the case, it could serve as a good rule of thumb to optimize seismic array design. Another important observation is the absence of maxima for $\bar{\lambda}\leq 3$ for the regular arrays, consistent with the fact that they suffer from spatial aliasing. In contrast, the SNR exhibits a clear maximum around the aliasing limit for the Square-Triangle and the Specter arrays. The Square-Triangle array performs better than the Specter array at lower $\bar{\lambda}$ than the Specter array, but its SNR remains low for $\bar{\lambda} \in [2, 3]$. 

This analysis shows that aperiodic tiles possess enhanced beamforming capabilities compared to regular arrays such as square or triangular lattice ones. Among the Tile($p$) family,  the window $p \in [0.45, 0.55]$ provides even better results by scattering the intensity of the first side lobes across a wide area and pushing the most energetic side lobes towards larger slowness or wavenumbers. The Specter array, in particular, shows more irregularities in the positions of its ARF side lobes than any other configuration (Fig.~\ref{fig:ARFSingleTile}\textbf{b}).
This provides more degrees of freedom to project a wavefield and boosts its beamforming capability. As the building blocks of aperiodic tilings, these tiles and metatiles naturally lack redundancy in their geometry, making them inherently more suitable for beamforming applications than regular lattices.

\section{Performance of multiple Tile($p$) seismic arrays and comparison with other aperiodic tilings
\label{sec:multipletile}
}

In the previous Section~\ref{sec:singletile}, we explored the array properties of individual Tile($p$) and compared them with regular or irregular arrays made of a similar number of stations. Here, we explore large-N arrays made of hundreds of stations. 

\begin{figure*}
    \begin{center}
    \includegraphics[width=0.99\textwidth]{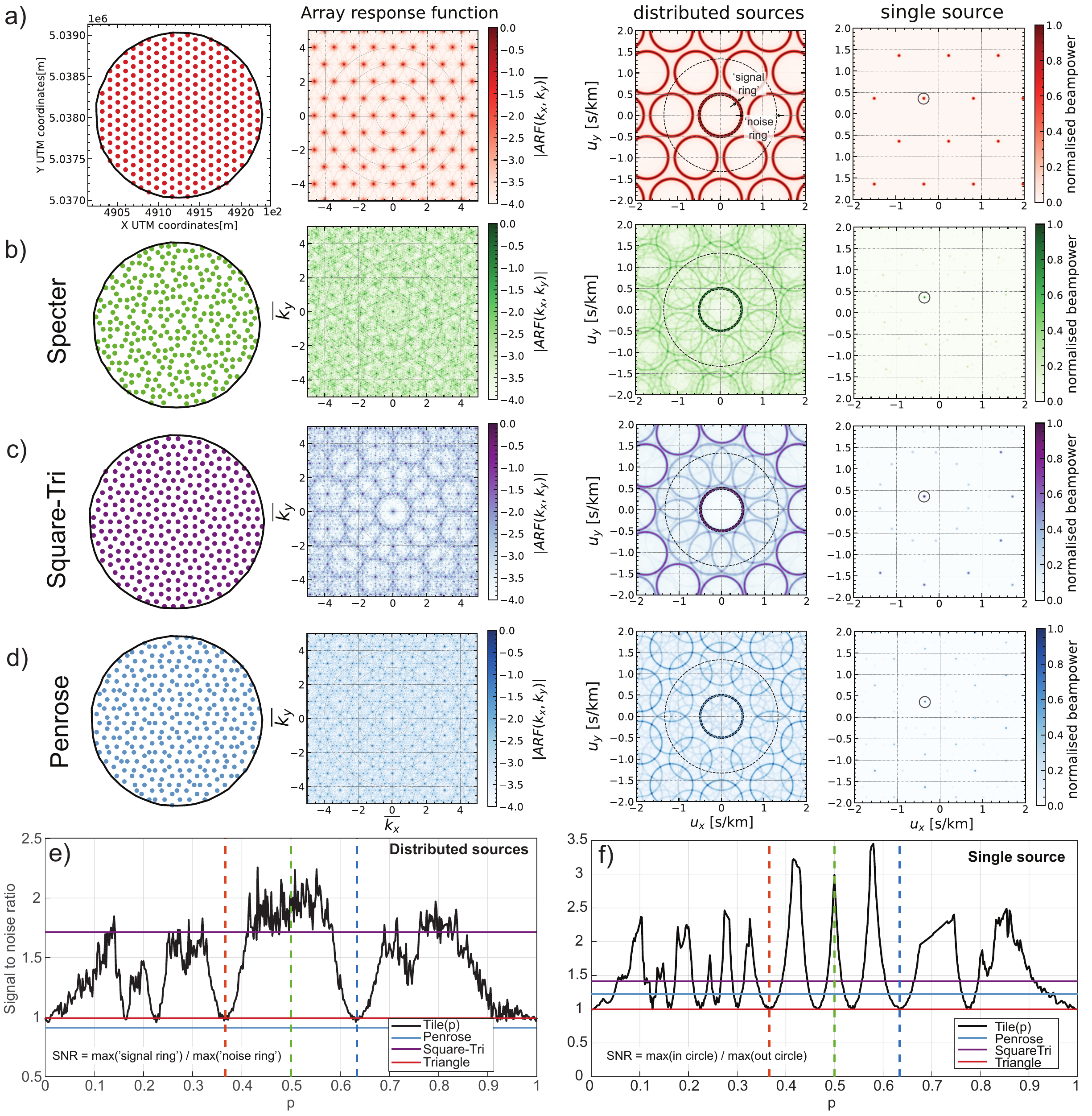} 
    \end{center}
    \caption{
    \textbf{Beamforming performances of Large-N aperiodic arrays}. a) from left to right: Station map of the triangular array, ARF (intensity in log scale), beamforming for distributed sources, beamforming for a single source. The rest of the columns show the same as in a) for the b) Specter array, c) the Square-Triangle array, and d) the Penrose array. The black circles in the distributed sources beamforming panels (third column) delineate the areas used to compute the SNR. The black circle in the single source beamforming panels delineates the areas used to compute the SNR. e) SNR of the distributed sources beamforming for the Tile($p$) arrays (black curve) as a function of $p$. The colored horizontal lines show the SNR level for the triangular, the Square-Triangle, and the Penrose arrays. The vertical dashed lines show the $p$ values of the Hat, the Specter, and the Turtle arrays. f) Same as e) for the single source beamforming results. In e) and f), SNR $\leq$ 1 when the signal is aliased and larger if not.
    }
    \label{fig:LargeNBeamformings}
\end{figure*}

To compare the beamforming capabilities of large-N Tile($p$)-based arrays and other aperiodic tilings, we generate arrays consisting of 310 stations (see Fig.~\ref{fig:LargeNBeamformings}).
The code accompanying this work can be used to generate seismic arrays of $N$ stations within a geographical region of choice, based on any Tile($p$), Square-Triangle, or Penrose tilings (see Fig.~\ref{fig:MakingLargeN} and Methods section).

For the following analysis, we choose a circular region, with an aperture of 2 km, see Fig.~\ref{fig:LargeNBeamformings}. 
On par with single-tile arrays (Fig.~\ref{fig:BeamformingSingleTile}), we simulate a wavefield generated by either distributed sources or a single source.
For each array, we fix the wavelength to be equal to twice the average nearest-neighbor distance (the aliasing limit) and a velocity of 2 km/s. 

The results are shown in Fig.~\ref{fig:LargeNBeamformings}, one line per array type:  Fig.~\ref{fig:LargeNBeamformings}\textbf{a} for the triangular array, Fig.~\ref{fig:LargeNBeamformings}\textbf{b} for the Specter array, Fig.~\ref{fig:LargeNBeamformings}\textbf{c} for the Square-Triangle array and Fig.~\ref{fig:LargeNBeamformings}\textbf{d} for the Penrose array. The first column shows the map of each array on the same scale, the second column is the ARF of each array with a logarithmic scale intensity, and the third and fourth columns are the beamforming results for the distributed and single-source scenarios, respectively. 

Figs.~\ref{fig:LargeNBeamformings}\textbf{e} and \textbf{f} display the SNR of the beamforming as a function of $p$ for the distributed sources and single source configurations, respectively. The horizontal lines in Fig.~\ref{fig:LargeNBeamformings}\textbf{e} and \textbf{f} indicate the SNR for the three other geometries: an aperiodic Penrose tiling in blue, an aperiodic Square-Triangle tiling in purple, and a regular triangular lattice array in red. 
We use the same SNR definition as for the single-tile beamforming (Fig.~\ref{fig:BeamformingSingleTile}) for both the distributed sources and single-source scenarios: the SNR is defined as the ratio of the maximum intensity within the 'signal ring' (the small black circles in Fig.~\ref{fig:LargeNBeamformings}, for the single-source scenario, and the inner ring for the distributed sources scenario) 
over the maximum intensity of the 'noise ring' (the rest of the diagram, out of the black circle for the single-source scenario, and the outer ring for the distributed sources scenario, see Fig.~\ref{fig:LargeNBeamformings}). With this definition, SNR $\leq$ 1 indicates that the signal is aliased. 

Overall, we see that aperiodic arrays perform better than regular ones. The SNR profile of large-N Tile($p$) for distributed sources, Fig.~\ref{fig:LargeNBeamformings}\textbf{e}, resembles that of single Tile($p$) arrays (Fig.~\ref{fig:BeamformingSingleTile}\textbf{f}). However, in the present case, the SNR level is higher than one across most $p$ values except for evident SNR reductions at both ends of the $p$ spectrum and four dips: two around $p\simeq0.16, 2.2$, and two for the Hat and the Turtle tiles (see orange and blue vertical dashed lines in Fig.~\ref{fig:LargeNBeamformings}\textbf{e}). These SNR reductions are again explained by the proximity of these tiles to a regular triangular lattice. In contrast to single tile arrays, the SNR maximum for $p \in [0.45, 0.55]$ is less pronounced. At this scale, the SNR of the Square-Triangle array is on the same level as the baseline SNR level of the Tile($p$) arrays for $p \in [0.45, 0.55]$. The Penrose tiling exhibits an SNR below one, indicating that the interfering aliased rings sum up to intensity values similar to the maximum intensity in the 'signal ring'. 

The single-source scenario, Fig.~\ref{fig:LargeNBeamformings}\textbf{f}, highlights other noteworthy features. First, the SNR is more spiky, and eleven discrete $p$ values produce Tile($p$) arrays that are entirely aliased. Among them, we find the Chevron ($p$=0), the Hat, the Turtle, and the Comet ($p$=1). The number of $p$ values outperforming (SNR $\geq$ 2.5) is also finite, peaking around three $p$ values showing a clearly defined maximum. The Specter is one of them. With this SNR definition, the Penrose and Square-Triangle arrays exhibit better performances than regular arrays, but they are far from the performances of the Specter array. For the distributed sources scenario, we tested another definition of the SNR: the Root-Mean-Squared (RMS) intensity ratio in the 'signal ring' and the 'noise ring' (see Appendix \ref{app:SNR}, Fig.~\ref{fig:SUPPL_SNRrmsrms}). Interestingly, whatever metric we choose to measure the SNR with, all aperiodic tilings outperform compared to the triangular case.

To test the robustness of the aperiodic arrays to sensor site perturbations, compared to regular arrays, we analyzed the effects of 1) disordering the sensors' positions and 2) the impact of sensor failure. To do so, we look at the beamforming SNR for distributed sources as a function of the type of tiling. In the first case, we randomly perturb the sensor positions, drawing 50 disordered realizations from a normal distribution centered on the actual sensor position with a variable standard deviation (0\% to 35\% of the average nearest-neighbor distance). In the second case, we remove a percentage of randomly chosen sites (repeating this operation 50 times to draw robust statistics). We remove from 1\% to 85\% of the sites. 

\begin{figure*}
    \begin{center}
    \includegraphics[width=\linewidth]{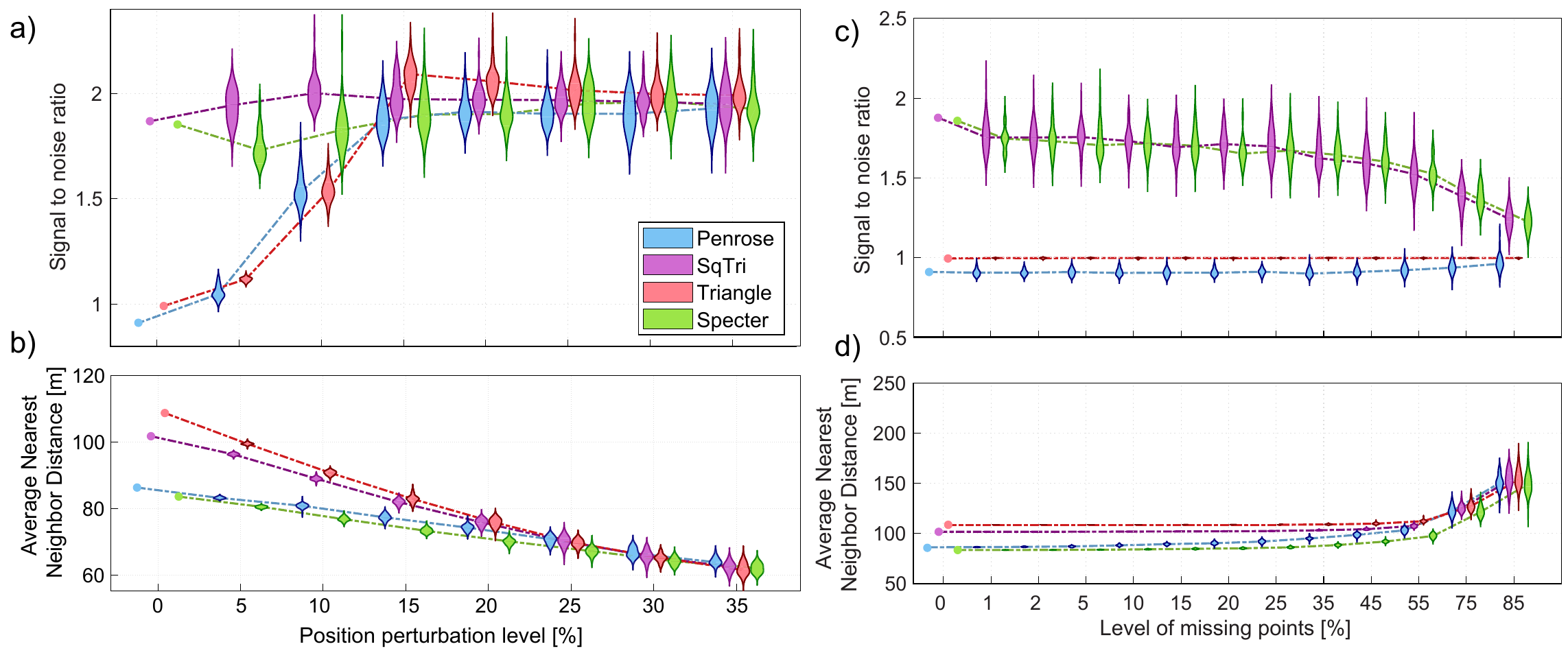} 
    \end{center}
    \caption{
    \textbf{Robustness to sensor misplacement and failure}. a) Analysis of the influence of random sensor misplacement on the beamforming SNR of four types of arrays as a function of the level of randomness (in percent of the average nearest-neighbor station distance in the unperturbed arrays). The beamforming is computed at the aliasing limit of the unperturbed arrays. b) Evolution of the average nearest-neighbor station distance as a function of the position perturbation level. c) Analysis of the influence of missing sensors on the beamforming SNR of four types of arrays as a function of the level of missing sensors (in percent of 310 sensors). The beamforming is computed at the aliasing limit of the unperturbed arrays. d) Evolution of the average nearest-neighbor station distance as a function of the level of missing sensors.  All points are calculated averaging over 50 disorder realizations.
    }
    \label{fig:LargeN_SNRvsNoise}
\end{figure*}

The results for the sensors' misplacement are displayed in Fig.~\ref{fig:LargeN_SNRvsNoise}\textbf{a}. We see that the beamforming results of the Specter and Square-Triangle arrays are unaffected by sensor misplacement. For the Triangle and Penrose arrays, increasing the level of misplacement increases the SNR. This is true up to a plateau where, with more than 15\% perturbation, the four types of arrays behave similarly. The Specter and Square-Triangle arrays behave similarly despite having different general sensor placements. A small level of randomness in the sensor positions improves the performance of regular arrays. 

The improvement of SNR with an increased randomization of the station positions can be due to two factors: 1) the overall reduction of the average nearest-neighbor station distance (Fig.~\ref{fig:LargeN_SNRvsNoise}\textbf{b}) as a function of the randomization level while the wavelength is fixed; 2) a better spatial sampling of the wavefield with fewer redundant elements in the arrays when the randomization is increased, or a combination of both effects. Given the linear trends of the average nearest-neighbor station distance curves (Fig.~\ref{fig:LargeN_SNRvsNoise}\textbf{b}) and the non-linear increase of SNR (Fig.~\ref{fig:LargeN_SNRvsNoise}\textbf{a}), we infer that the reduction of the average nearest-neighbor station distance is not the sole responsible for the improvement of the SNR but that the reduction of array redundancies plays a dominant role.

From the average nearest-neighbor station distance, we also see that the Specter and Penrose arrays have a smaller average nearest-neighbor station distance for the same spatial support and number of stations. This means that for a given wavelength, we would expect these two geometries to perform better with respect to aliasing as $\bar{\lambda}$ is increased. 
However, as discussed in the next section, this is not the case for the Penrose tiling, whose side lobes are not only controlled by its interstation distance but also by its azimuth distribution, resulting in a low SNR (see Section \ref{sec:Disc}).

The results for the sensor failures are displayed in Fig.~\ref{fig:LargeN_SNRvsNoise}\textbf{c}. Again, the Specter and Square-Triangle arrays outperform the two other geometries with systematically larger SNR values. The reduction of the number of sensors reduces the SNR, and the reduction seems directly correlated with the increase of the average nearest neighbor distances with each array (Fig.~\ref{fig:LargeN_SNRvsNoise}\textbf{d}). It is worth noticing that for 310-station arrays, the random failure of up to one quarter of the sensors has little effect on their beamforming performance. With the RMS definition (Fig.~\ref{fig:SUPPL_SNRrmsrms}\textbf{c}), we see that it is only after removing more than 50\% of the stations that the aperiodic arrays reach the SNR level of the periodic one, indicating the high level of redundancy of the aperiodic arrays.

\section{Discussions and Conclusion \label{sec:Disc}}

We have explored a newly discovered Hat family of monotile aperiodic tilings \citep{smith2023aperiodic, smith2023chiral} from the viewpoint of (seismic) array analysis. The comparison with other well-known aperiodic tilings, such as the Penrose and Square-Triangle tilings, shows that the Tile($p$) tiling could be an ideal geometry for some range of $p$ values. MAS arrays made of one Tile($p$), i.e., 14 stations, exhibit enhanced beamforming performances with larger SNR values for $p \in [0.45, 0.55]$ (Fig.~\ref{fig:BeamformingSingleTile}) for configurations with distributed sources around the array. For MAS arrays made of hundreds of stations, most of the $p$ spectrum produces large SNRs except for a few notable exceptions: the extreme $p$ values and the Hat and the Turtle (Fig.~\ref{fig:LargeNBeamformings}\textbf{e}). For the single source configuration (Fig.~\ref{fig:LargeNBeamformings}\textbf{f}), we obtain more discrete outperforming $p$ values, which restrict to $p \in [0.41, 0.43] \cup [0.495, 0.505] \cup [0.57, 0.59]$. For these ranges, the side lobes' amplitude is more than 2.5 times smaller than the amplitude of the main lobe, which efficiently mitigates aliasing problems. 


An important transversal question for all our analyses is how to compare each geometry fairly. For the single-tile analysis (Fig.~\ref{fig:ARFSingleTile}), we displayed the ARFs normalized by the smallest inter-station distance in the array. This is a standard approach; however, for Tile($p \in [0.45, 0.55]$), this minimum distance is not one of the sides of the tile but the distance at the "neck" of the Specter. For this range of $p$, the minimum distance is unique, representing only about 1\% of all possible distances in the array, whereas, for other $p$-values, the minimum distance is one of the sides and is present multiple times. Therefore, normalizing by the minimum distance might present the Specter and other central-$p$ tiles slightly more favorably than they actually are. On the other hand, this approach allows us to compare all types of arrays consistently; this is why we favor it.

Another aspect of quantifying the beamforming performance that relies on a subjective choice is the definition of the SNR. Here, we use the ratio of maximum intensity values between the area of interest of the beamforming diagram and an area that we define as 'noise'. This definition is a conservative proxy for aliasing but cannot describe more subtle features of the beamforming diagrams, especially for the distributed sources scenarios. For these cases, we tested another definition of the SNR (see Appendix \ref{app:SNR}): the ratio between the Root-Mean-Squared (RMS) beamforming intensity of the 'signal ring' and the 'noise ring.' This definition measures how well the intensity is scattered and attenuated in the 'noise ring' but will fail to clearly indicate if the wavefield is aliased or not. This is why we favor the maximum ratio definition in the main text. Figure~\ref{fig:SUPPL_SNRrmsrms} shows the SNR results using the RMS definition for the different tests involving the distributed sources scenario. With this definition, the Square-Triangle geometry outperforms the other geometries. Still, the central portion of the $p$ spectrum exhibits an SNR maximum. 

When comparing a single Tile($p$) to other geometries (triangular and Square-Triangle tiling, Fig~\ref{fig:BeamformingSingleTile}), we had to reach a compromise between the spatial support,  the number of stations, and the minimum inter-station distance. It was impossible to set these three parameters equal to the ones of the Tile($p$) simultaneously. The same principle applies to the comparisons in the large-N array cases. We chose to fix the number of stations and the spatial support rather than fixing the minimum interstation distance and varying the number of stations. It is a scenario closer to the actual seismological practice of array deployment, where we usually have access to a limited number of stations.  

\begin{figure*}
    \begin{center}
    \includegraphics[width=\linewidth]{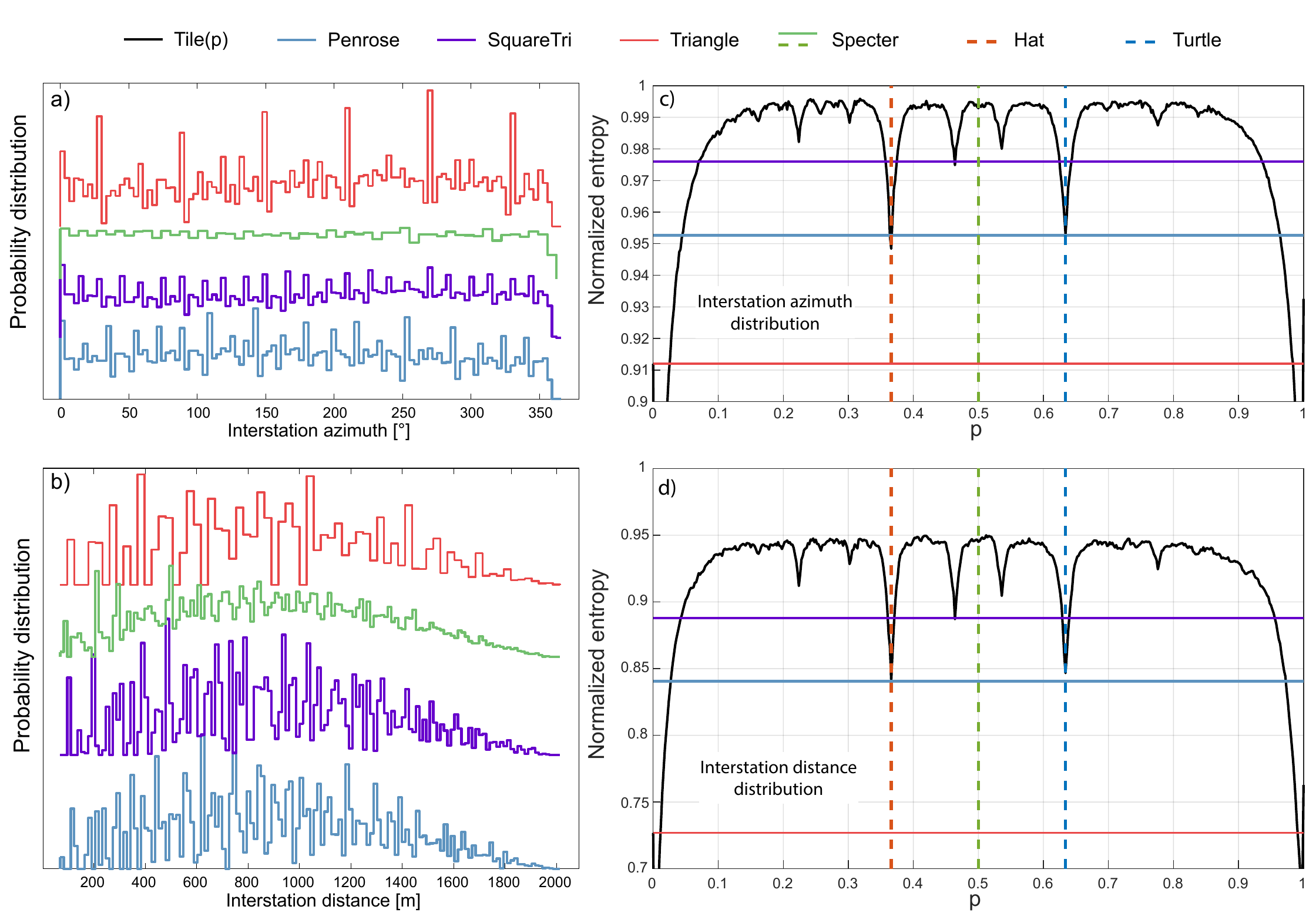} 
    \end{center}
    \caption{ \textbf{Tile($p$) interstation azimuth and distance distributions entropy.} Distributions of interstation azimuths (a) and distances (b) for the four 310-station arrays used in Section \ref{sec:multipletile}. The normalized entropy (see Eq.~\eqref{eq:entropy}) of each distribution is shown by the horizontal lines in c) and d) as well as the normalized entropy of 310-stations Tile($p$) arrays with the black curves. The Specter array has the most homogeneous distributions, with a normalized entropy closest to 1. 
    }
    \label{fig:Entropy}
\end{figure*}

Our analysis, spanning several scenarios, while not exhaustive (we, for example, did not test the effect of noisy seismic data, only noisy station positions or missing sensors), is comprehensive enough to demonstrate that Specter arrays (and two other geometries that we conjecture to be around $p = \sqrt{2}-1 \approx 0.42$ and $p = 2 -\sqrt{2} \approx 0.58 $) perform better than regular arrays and some aperiodic ones in general. 

The aperiodic nature of the Specter tiling is not the fundamental reason it performs so well: The Hat and Turtle tilings are clear counterexamples of aperiodic tiling suffering from the same aliasing limits as regular arrays. Nor is its monotile nature: the Square-Triangle tiling, made with two tiles, performs almost as well as the Specter tiling. The Specter monotile and large-N arrays made of it are ideal geometries for beamforming applications because they sample azimuths and distances more evenly than any other tested geometry. This is illustrated in Fig.~\ref{fig:Entropy}\textbf{a}-\textbf{b}, where we can see that the distribution of interstation azimuths is almost uniform for the Specter array. 

In contrast, other geometries have oversampled and undersampled azimuthal directions. This is particularly evident for the regular array. The interstation distance distribution is much smoother for the Specter array than for the other arrays. We confirmed these observations quantitatively by computing the normalized entropy of each distribution (Fig.~\ref{fig:Entropy}\textbf{c}-\textbf{d}), which estimates the distance of a distribution from a pure uniform distribution: 
\begin{equation}
\label{eq:entropy}
      \bar{E} =  -\frac{\sum_{i = 1}^{n} p_i \log2(p_i)}{\log2(n)},
\end{equation}
where $p_i$ is the normalized histogram count of the observable, here either the interstation azimuth or the interstation distance, in the bin $i$ and $n=361$ is the number of bins.
%
%
Normalized entropy values closer to 1 indicate proximity to a uniform distribution. Again, we observe that most Tile($p$) exhibit a large entropy, except for a few $p$ values, which means that in general, Tile($p$) provides a very uniform distribution of interstation azimuths and distances (Fig.~\ref{fig:Entropy}\textbf{c}-\textbf{d}). This uniform azimuthal and distance sampling provides the largest number of degrees of freedom possible to sample a wavefield, equivalent to having a coarray (see Methods) with its sites as homogeneously distributed as possible. 

Achieving a homogeneous distribution of points is also possible by drawing coordinates from a uniform distribution. However, homogeneity is guaranteed only on average. In realistic scenarios, only a single realization will be deployed, which is not guaranteed to have uniform azimuthal and distance sampling. In contrast, the homogeneity of the Tile($p$) family is not an average property, and thus, wave sampling will be reproducible and deterministic in practice, unlike placing stations at random. 

Beyond beamforming applications, the ability to homogeneously sample space can be precious for seismic imaging applications, particularly ambient seismic noise tomography using large-N arrays \citep[e.g.,][]{lin2013high,mordret2013near,mordret2019shallow,chmiel2019ambient, Tsarsitalidou2024}. In this technique, homogenizing the rays' azimuthal distribution will produce velocity models where anomalies retain their shape and are less likely smeared by the array imprint, improving their resolution and fidelity. Historically, researchers have used specific and different array designs for beamforming and imaging. Spiral arrays are ideal for beamforming due to their optimal distance and azimuth sampling \citep{kennett2015spiral,sheng2024tracking}. However, the ray coverage they produce is not well suited for imaging. On the contrary, dense arrays designed for imaging are often regular \citep[e.g.,][]{mordret2013near,sheng2022monitoring} and may suffer from aliasing when the wavelength becomes smaller than twice the average nearest-neighbor interstation distance. From this point of view, MAS Specter arrays are more versatile and optimal for beamforming and imaging applications altogether.

Our results can also be tested in settings beyond seismology, in physical realizations where the ARF or structure factor can be measured directly. 
One of these platforms is 3D-printed or nano-patterned disorder photonic crystals. 
In this context, exploring the Hat family as a design principle for photonic applications, such as band-gap optimization or cloaking~\citep{Vynck2023} is worth exploring. 
Similar arguments apply to other metamaterial platforms, such as acoustic arrays~\citep{Cheron2022, Tang2023}.
Another platform properties where our results can be tested are molecular patterns of the surface of a metal like copper using carbon dioxide molecules~\citep{GKW12}.
This platform has already been used to design Penrose tilings, where the structure factor was directly measured~\citep{Collins2017}.
Thus, it is a promising system to realize the Tile($p$) tiling and test our findings.
Similarly, ultra-cold atomic platforms~\cite{Bloch2008} realizing the Hat tiling could benefit from the results presented here.
For example, particle propagation can show characteristic phenomenology in aperiodic lattices~\citep{Sbroscia2020}.

Lastly, as discussed in the introduction, beamforming analysis is strongly tied to the diffraction pattern analysis of quasicrystals. 
In this regard, the quasi 12-fold symmetry of the Specter tiling stands out, as it should be explored further to determine the relationship, if any, with other 12-fold symmetry aperiodic tilings, such as the Square-Triangle tiling. 
Although these further analyses are beyond the scope of this study, we are confident that research on the Hat family of tiling, particularly the Specter tiling, is only in its infancy and will flourish soon. We suggest exploring fields and problems where alignments of points or structures, or lack thereof, are essential, as there is strong potential for practical applications in diverse areas.
Besides those mentioned above, we can include wind-farm optimization design or seismic cloaking to cite a few examples.

In conclusion, our work proposes an advantageous design principle based on the recently discovered aperiodic monotiles for seismic arrays to beat the Whittaker-Nyquist–Shannon (WSN) aliasing limit. Monotile aperiodic seismic (MAS) arrays outperform regular and other quasicrystalline arrays in signal-to-noise ratio for single and distributed source scenarios and demonstrate robustness to station position noise, providing a reliable solution in seismic array design. 
The array-response-function and beamforming analysis we presented here can be directly applied, without any conceptual modification, to rationalize generic wave scattering off aperiodic monotile sites. Hence, the benefits of MAS arrays can be directly exported to numerous other fields, e.g., the placement and design of telecommunications antennae, ad-atoms in solid-state physics, and compressed sensing of various digital signals.

\section*{Acknowledgments}

We thank P. Roux and L. Seydoux for discussing and suggesting randomizing the station positions. We thank the two reviewers for their suggestions to test the effects of removing stations and for their comments, which helped improve this paper. We are also grateful to C. de France for the hospitality and G. Hendrick for the discussions while starting this project.
We thank F. Flicker, S. Franca, H. Roche, M. Yoshii, and J. Schirmann for discussions and related collaborations. 
A. G. G. acknowledges support from the European Research Council (ERC) Consolidator grant under grant agreement No. 101042707 (TOPOMORPH). A. M. acknowledges support from the Agence Nationale pour la Recherche (ANR) through the project ANR-22-CE49-0019 (MACIV). 

\section*{Data and code availability}
The Python Jupyter Notebooks used to produce the tiling data and figures are available on the GEUS Dataverse with the following DOI: \url{https://doi.org/10.22008/FK2/G06YCK}.

\section*{Competing interests}
The authors have no competing interests.

\section*{Authors contributions}
Conceptualization: Aurelien Mordret and Adolfo G. Grushin; 
methodology: Aurelien Mordret;
software: Aurelien Mordret and Adolfo G. Grushin;
validation: Aurelien Mordret and Adolfo G. Grushin;
formal analysis: Aurelien Mordret and Adolfo G. Grushin;
writing original draft, review, and editing: Aurelien Mordret and Adolfo G. Grushin;




\appendix

\section{Plane-wave beamforming}

\begin{figure}[ht!]
    \begin{center}
    \includegraphics[width=\linewidth]{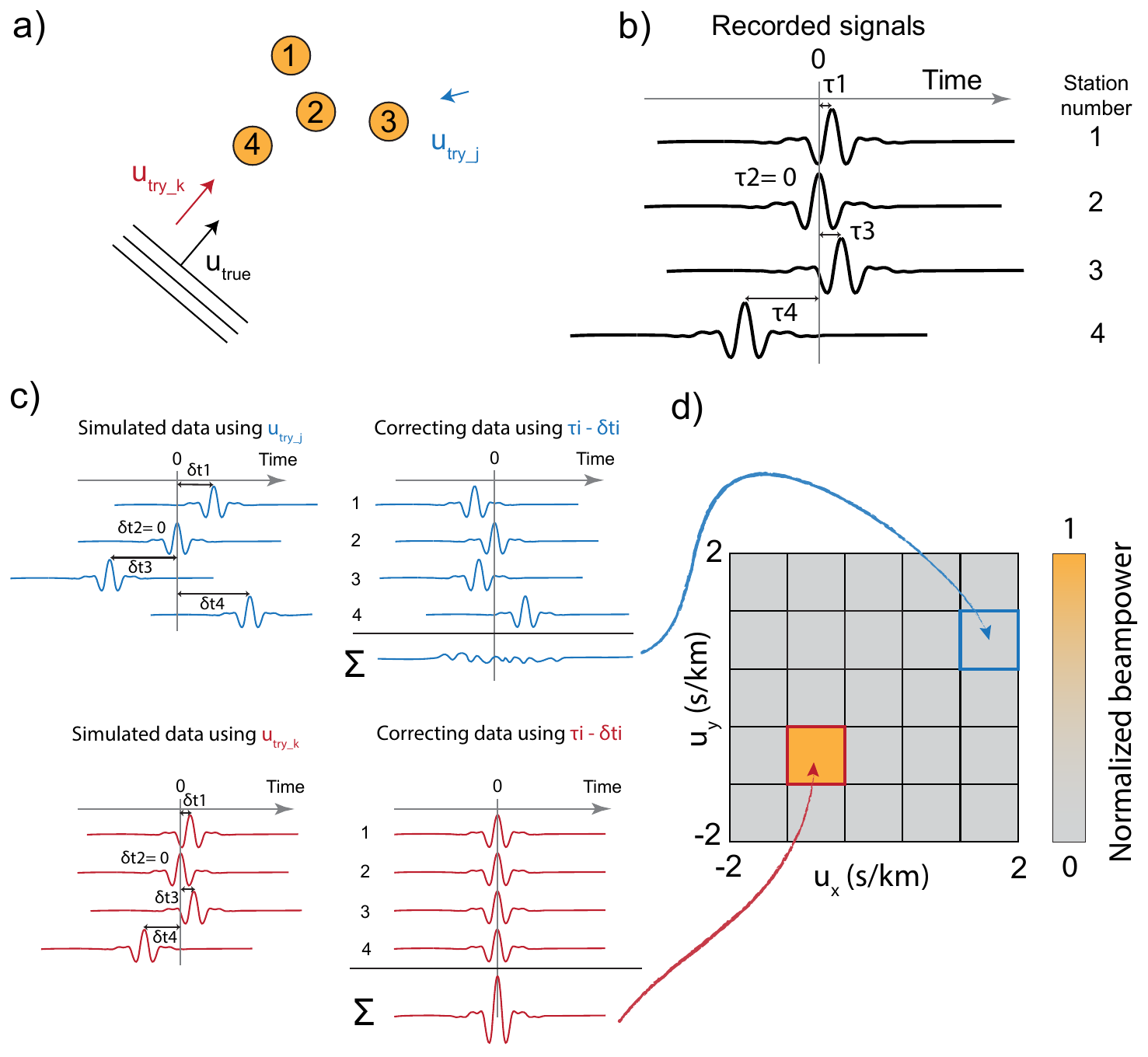} 
    \end{center}
    \caption{
    \textbf{Beamforming principle}. a) Map of the seismic arrays. The arrows show the (unknown) true slowness vector of the incoming plane wave (black) and two trial slowness vectors in red and blue. b) The signal recorded by the four seismic stations, with station 2 as the reference station. c) Correction of delays and sum of the signals according to the trial slowness vectors. d) The intensities of the stacked signals are displayed on the beamforming diagram at the positions of the trial slowness vectors (one per pixel).  The intensity is maximum for a slowness trial vector close to the true slowness vector.  
    \label{fig:BeamformingSUPPL}}
\end{figure}

Plane-wave beamforming is used to determine the unknown slowness parameters of waves recorded by a seismic array. The principle of plane-wave beamforming is illustrated in Figure~\ref{fig:BeamformingSUPPL}. The signal data recorded by the array are shifted in time according to trial slowness vector parameters and then stacked. The maximum intensity of the stacked signals is displayed on a matrix at the position of the trial slowness vector. A grid search is performed over each pixel of the beamforming diagram. The beamforming diagram exhibits a maximum intensity for a slowness vector matching the true slowness of the plane wave recorded by the array. 

Mathematically, beamforming can be expressed as follows. We will describe an array of $N$ stations that record a seismic signal 
$\mathbf{s}(\mathbf{x}_i, t) = \mathbf{s_i}(t)$, where $i = [1, ..., N]$ is the station index, $\mathbf{x}_i = [x_x, x_y, x_z]$ the geographical coordinates of the station $i$ with respect to the centroid $x_0$ of the array, and $t$ is the time. We are interested in planar arrays, so we assume $\mathbf{x}_i = [x_x, x_y]$. A plane-wave $v(\mathbf{k_0}) = e^{j\mathbf{k_0}\cdot \mathbf{x}}$, with $j=\sqrt{-1}$, traveling across the array with a wavenumber vector $\mathbf{k_0} = [k_x, k_y, k_z]$, will arrive at each station with a time-delay $\mathbf{\tau}_i=\mathbf{x}_i \cdot \mathbf{k_0} = \mathbf{x}_i \cdot (\omega \cdot \mathbf{u_0} )$. Here, $\omega = 2\pi f$ is the angular frequency, $f$ is the frequency, and $\mathbf{u_0}$ is the slowness vector, i.e., the inverse of the plane-wave velocity. The array-stacked signal, the sum of all signals, is maximum when all the traces from each station are in phase. This happens when the time-delay $\mathbf{\tau}_i$ is adequately corrected for at each station before summation, i.e., when the wavenumber used to compute the time-delay is equal to the actual wavenumber of the incoming plane wave. 
%

The beampower or beamforming intensity (abbreviated to beamforming in this work) $B(\mathbf{k}, \omega)$, i.e., the intensity
of the array-stacked signal, 
computed for trial plane-waves with wave vectors $\mathbf{k}$ and at frequency $\omega$ is \citep{gal2019beamforming}

\begin{equation}
    B(\mathbf{k}, \omega) = \frac{1}{N^2}\left| \sum_{i=1}^N s_i( \omega) e^{j \mathbf{x}_i \cdot \mathbf{k}}  \right|^2. 
    \label{eq:beam}
\end{equation}
In Eq.~\eqref{eq:beam}, $\mathbf{s}_i( \omega)$ is the temporal Fourier transform of $\mathbf{s}_i( t)$ defined by

\begin{equation}
    \mathbf{s_i}( \omega) = \int^{\infty}_{-\infty} \mathbf{s}_i( t) e^{j \omega t} dt.
\end{equation}

For horizontally propagating waves, the wave vector simplifies to $\mathbf{k} = [k_x,k_y] = \omega \cdot \mathbf{u}$, where $\mathbf{u}= [u_x,u_y]$ is the horizontal slowness vector. The beamforming is maximized for wavenumbers corresponding to actual plane waves crossing the array from the right azimuths with the right slowness. 

Alternatively, the beamforming can be written as

\begin{equation}
  \begin{aligned}
        B(\mathbf{k}, \omega) &= \frac{1}{N^2}\left( \sum_{i=1}^N s_i( \omega) e^{j \mathbf{x_i} \cdot \mathbf{k}}  \right)^* \times \left( \sum_{l=1}^N s_l(\omega) e^{j \mathbf{x}_l \cdot \mathbf{k}}  \right), \\
    & = \frac{1}{N^2} \left( \sum_{i=1}^N \sum_{l=1}^N e^{-j \mathbf{x}_i \cdot \mathbf{k}} \mathbf{C}_{il}(\omega) e^{j \mathbf{x}_l \cdot \mathbf{k}} \right),\\
    & = \frac{1}{N^2} \left( \sum_{i=1}^N \sum_{l=1}^N \mathbf{C}_{il}(\omega) e^{j \mathbf{r}_{il} \cdot \mathbf{k}} \right),
    \label{eq:beam2}
\end{aligned}
\end{equation}
where $\mathbf{C}_{il} = s_i( \omega)^* s_l(\omega)$ is the covariance matrix of the data \citep[e.g.,][]{seydoux2016detecting}
, $\mathbf{r}_{il}=\mathbf{x}_l-\mathbf{x}_i$ is the inter-station distance between sensors $i$ and $l$ 
, and $^*$ is the complex conjugate \citep{gal2019beamforming}. Therefore, plane-wave beamforming can be seen either as the projection of the physical array data onto plane waves (Eq.~\eqref{eq:beam}) or the projection of the physical array data cross-correlations (Eq.~\eqref{eq:beam2}), i.e., the virtual data from a virtual array made of $N^2$ stations located at coordinates $\mathbf{r}_{il}$, projected onto plane-waves. This virtual array is called the \textit{difference coarray} \citep{hoctor1990unifying, hopperstad98_norsig}, or simply \textit{coarray} in this work, and has been introduced in array seismology by \cite{haubrich1968array}. The coarray is the spatial auto-correlation of the physical array coordinates and is defined as the set $\mathbb{A}$:

\begin{equation}
    \mathbb{A} = \left\{ \mathbf{y} | \mathbf{y} = x_2 - x_1, \forall x_1, x_2 \in \mathbf{X} \right\},
\end{equation}
where $\mathbf{X}$ is the set of coordinates of the physical array. In mathematical morphology \citep{serra2022mathematical}, the coarray is equivalent to the dilation of $\mathbf{X}$ by itself. The number of unique locations in the coarray and the homogeneity of their spatial distribution determine the array's resolution and anti-aliasing capabilities. 

\subsection{Array response function}

The array response function (ARF) is usually defined as the beamforming for a plane-wave illuminating all the stations of the array at the same time from below, i.e., a plane-wave with an infinite velocity or a wavenumber $\mathbf{k}=0$. From Equation~\eqref{eq:beam}, the ARF simplifies as

\begin{equation}
    \mathrm{ARF}(\mathbf{k}, \omega) = \frac{1}{N^2}\left| \sum_{i=1}^N e^{j \mathbf{x_i} \cdot \mathbf{k}}  \right|^2. 
    \label{eq:ARF}
\end{equation}

Similarly, Equation~\eqref{eq:beam2} gives

\begin{equation}
    \mathrm{ARF}(\mathbf{k}, \omega) = \frac{1}{N^2} \left( \sum_{i=1}^N \sum_{l=1}^N e^{j (\mathbf{x}_l -  \mathbf{x}_i) \cdot \mathbf{k}} \right),
    \label{eq:ARF2}
\end{equation}
which has the same support as the spatial Fourier transform of the coarray. Both the ARF and the coarray encapsulate the geometrical features of the physical array and its capabilities in terms of resolution and robustness to aliasing effects. In the [$k_x$, $k_y$] plane, the ARF exhibits a central lobe at $k_x$ = $k_y$ = 0 and several side-lobes more or less far from the main lobe with positions and amplitudes depending on the array geometry. In solid-state physics, the ARF  is called the structure factor, and it is defined by the atomic positions and measured by X-ray scattering. \citep[e.g.,][]{Levine84}.

\subsection{Resolution and aliasing}

The resolution of an array is determined by its aperture, i.e., the largest interstation distance within the array, $r_{max} =  \max\{\mathbf{r}\}$, and can be measured from the width of the main central lobe of the ARF. The resolution assesses how well a given array can distinguish two plane waves with similar slowness. A standard approximation of the slowness resolution is $ u_{R} = 1/(2 f r_{max}) $, using the frequency $f$. Inversely, aliasing is controlled by the smallest interstation distance in the array, $r_{min} =  \min\{\mathbf{r}\}$, and is manifested by the repetition of the ARF pattern at larger wavenumber or slowness, according to the WNS theorem. Based on regular array approximation, the distance in slowness at which the pattern repeats is given by $u_{WNS} = 1/(2 f r_{min}) $. The ideal array design will aim at narrowing the main lobe while reducing the amplitudes of the side lobes and pushing them towards higher slowness. 

\begin{figure}[ht!]
    \begin{center}
    \includegraphics[width=\linewidth]{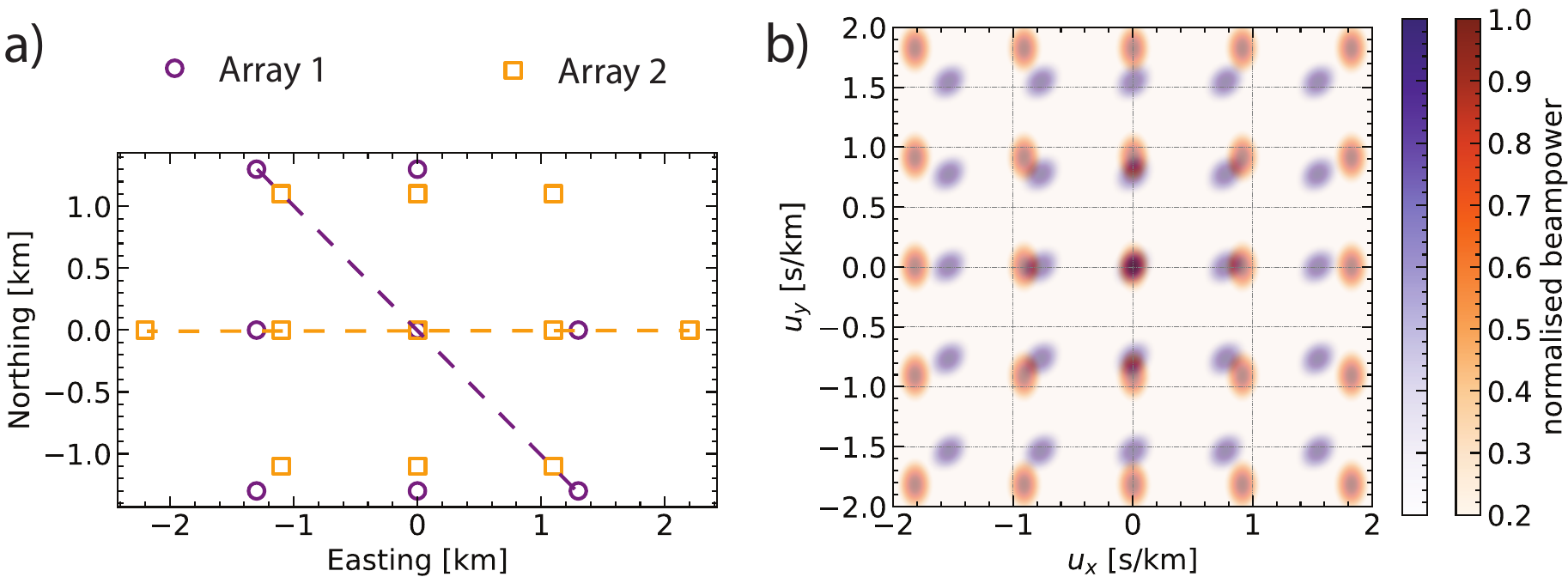} 
    \end{center}
    \caption{
    \textbf{Resolution and aliasing illustration for two regular arrays}. a) Map of the two arrays. The dashed lines show the two arrays' maximum inter-station distances or apertures. b) Superposition of the ARFs of the two arrays, in purple for Array 1 and in orange for Array 2.      
    \label{fig:ARFSupplementary}}
\end{figure}
Fig.~\ref{fig:ARFSupplementary} illustrates the effect of the array design on resolution and aliasing. Array 1 (purple circles in Fig.~\ref{fig:ARFSupplementary}\textbf{a} comprises eight stations located on a 1.1 km square lattice with an aperture of 3.11 km. Array 2 (orange squares in Fig.~\ref{fig:ARFSupplementary}\textbf{a}) comprises eleven stations assembled on a 1.3 km square lattice with a 5.2 km aperture. The corresponding apertures are indicated by dashed lines in Fig.~\ref{fig:ARFSupplementary} \textbf{a}. Fig.~\ref{fig:ARFSupplementary}\textbf{b} shows the ARF for both arrays. We can see how the lobes of Array 2 are narrower thanks to the larger aperture and are pushed away to higher slowness thanks to the smaller interstation distances. Note that the lobes for both arrays are elliptical because the two arrays are elongated in the North-West and East-West directions for Array 1 and 2, respectively.

\section{Building Tile($p$) arrays of N stations}


 In this section, we will show how to design large arrays made of multiple Tile($p$) given a user-defined number of stations and a predefined spatial support. We include a suite of Jupyter Notebooks (see Data and Code availability section) to generate such arrays. There are four Notebooks available for four different tilings: Achiral Tile($p$) (Fig.~\ref{fig:HatFamily}a), Chiral Specter (Fig.~\ref{fig:HatFamily}b), Square-Triangle (Fig.~\ref{fig:HatFamily}d), and Penrose-rhombic aperiodic tilings (Fig.~\ref{fig:HatFamily}c). The Penrose and Square-Triangle tiling generators are Python translations from Matlab codes by \citet{eddins2024penrose} and \citet{ho2024generate}, respectively. The Tile($p$) and Chiral Specter tilings are translated from \citet{weblink}. The remaining Notebooks in the supplementary package reproduce relevant parts of the figures displayed in this paper.

\begin{figure*}
    \begin{center}
    \includegraphics[width=\linewidth]{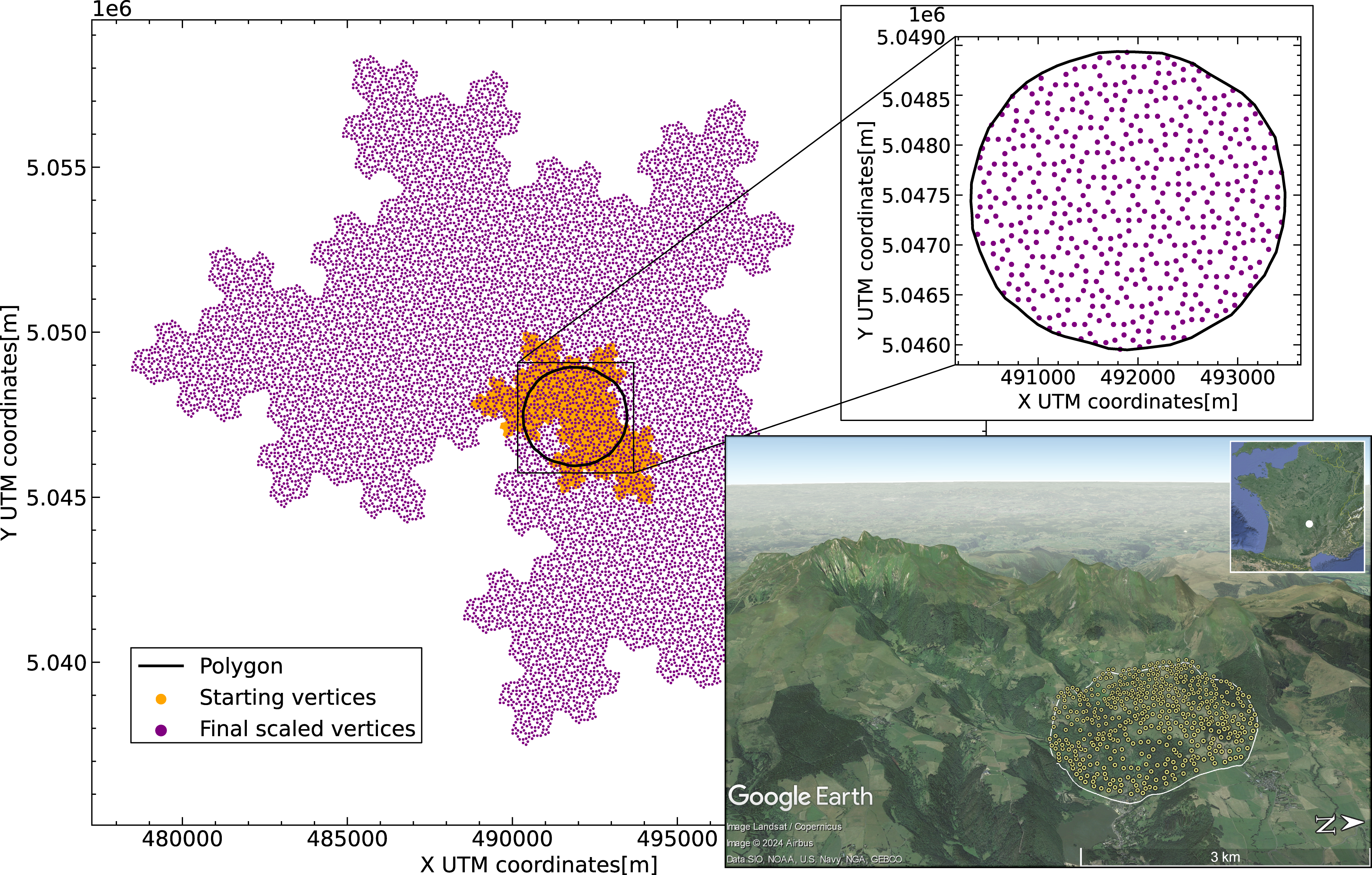} 
    \end{center}
    \caption{
    \textbf{Deployment simulation of Large-N Tile($p$) MAS array}. Generating a 499-station achiral Specter array within a circular area located in the French Massif Central. The Google Earth inset shows the location of the array wrapped on an exaggerated topography ($\times$2.5).
    }
    \label{fig:MakingLargeN}
\end{figure*}

The first step to generating a seismic array is to define its spatial support. Here, we use \citet{googleearth2024} to create a \textit{polygon} encompassing the area where one wants to install the seismic array (black contour in Fig.~\ref{fig:MakingLargeN}), and we save it as a KML file. 
The code reads the coordinates of the polygon in the Universal Transverse Mercator (UTM) format and requires that the polygon be defined on a unique UTM zone. The second step is to define the number of stations $N$ (plus or minus a small range $\delta N$, to allow the algorithm to converge more easily) that one wants to fit within the polygon. The third step is generating a large enough tiling with more (unique) vertices than the user-predefined number of stations (orange dots in Fig.~\ref{fig:MakingLargeN}). The algorithm will then scale the vertices iteratively with a user-defined scaling factor until $N \pm \delta N$ vertices fit within the polygon (purple dots in Fig~\ref{fig:MakingLargeN}). The selected vertices coordinates are then saved in a KML file. Fig.~\ref{fig:MakingLargeN} shows an example of designing a 499 stations array located in the French Massif Central within a circular domain based on an achiral Specter tiling. 

\section{Signal to noise results using a Root-Mean-Squared definition
\label{app:SNR}}

In the main text, we defined the signal-to-noise ratio (SNR) as the ratio of maximum intensity values between the area of interest of the beam-forming diagram and an area that we defined as ’noise.’ This definition is a conservative proxy for aliasing but cannot describe more subtle features of the beamforming diagrams, especially for the distributed sources scenarios. 

\begin{figure*}
    \centering
    \includegraphics[width=0.99\linewidth]{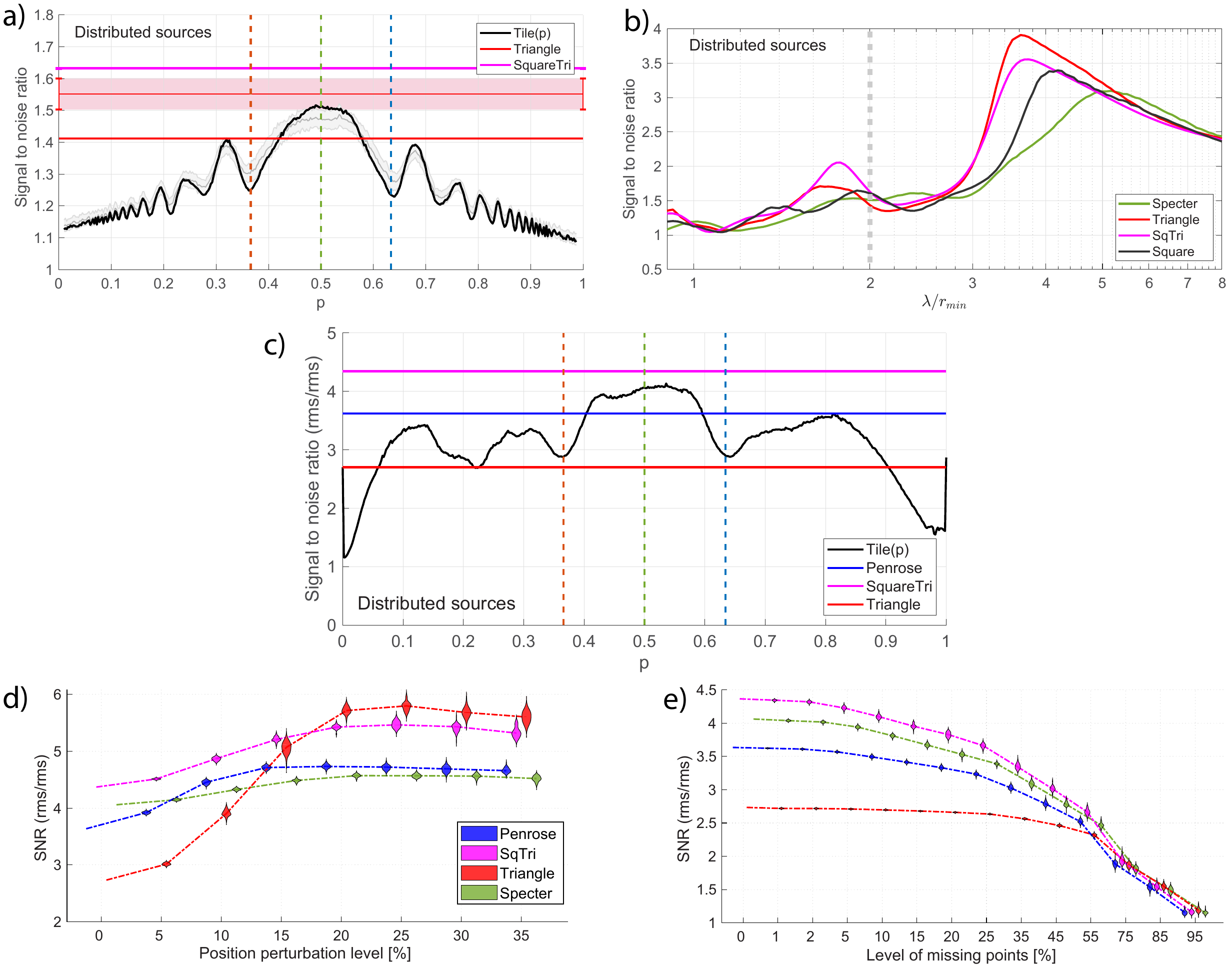}
    \caption{\textbf{Signal-to-Noise benchmarks using the Root-Mean-Squared criterion.} These figures contrast those discussed in the main text, which were computed using the maximum-ratio definition.  
    a) SNR of the beamforming results for the distributed sources as a function of the parameter p, to be compared with Fig.~\ref{fig:BeamformingSingleTile} f). 
    b) Beamforming SNR as a function of the
wavelength, to be compared with Fig.~\ref{fig:SNRvsLambda}.
c) SNR of the distributed sources
beamforming for the Tile($p$) arrays (black curve) as a function of $p$. The colored horizontal lines show the SNR level for the
triangular, the Square-Triangle, and the Penrose arrays. To be compared with Fig.~\ref{fig:LargeNBeamformings}e). 
d) Analysis of the influence of random sensor misplacement on the beamforming SNR of four types of arrays as a function of the level of randomness (in percent of the average nearest-neighbor station distance in the unperturbed arrays). To be compared with Fig.~\ref{fig:LargeN_SNRvsNoise}a). 
e) Evolution of the
average nearest-neighbor station distance as a function of the level of missing sensors. To be compared with Fig.~\ref{fig:LargeN_SNRvsNoise}c).}
    \label{fig:SUPPL_SNRrmsrms}
\end{figure*}

In this Appendix, we present the results for comparison using an alternative definition of the SNR: the ratio between the Root-Mean-Squared (RMS) beamforming intensity of the ’signal ring’ and the ’noise ring’. This definition measures how well the intensity is scattered and attenuated in the 'noise ring' but will fail to clearly indicate if the wavefield is aliased or not. This is why we favor the maximum ratio definition in the main text. 

Figure~\ref{fig:SUPPL_SNRrmsrms} shows the SNR benchmarks discussed in the main text computed here using the RMS definition for the different tests in the main text involving the distributed sources scenario. With this definition, the Square-Triangle geometry generally outperforms other geometries. Still, the central portion of the $p$ spectrum exhibits an SNR maximum, and the Hat and Turtle tilings show local SNR minima (Fig.~\ref{fig:SUPPL_SNRrmsrms}a-c). The SNR as a function of the normalized wavelength (Fig.~\ref{fig:SUPPL_SNRrmsrms}b) shows again a sharp increase after $\bar{\lambda} = 3$ and a local maximum before $\bar{\lambda} = 2$ for all arrays except the Specter geometry. This shows that the RMS metric is not appropriate for properly conveying the quality of a beamforming diagram. In this case, strongly aliased signals and particular geometries can produce interferences in beamforming diagrams and reduce the RMS intensity in the 'noise ring'. Interpreting these diagrams regarding sources' azimuthal distribution and slowness should be done cautiously. Overall, the Specter geometry produces small SNR values with the RMS definition compared to the other geometries because the Specter tiling has a very scattered, almost non-discrete (continuous) Fourier decomposition (Fig.~\ref{fig:LargeNBeamformings}), with a non-zero background intensity.

\newpage

\bibliography{sn-bibliography}
\end{document}